\documentclass[amsmath,superscriptaddress,aps,pra,twocolumn]{revtex4-2}

\usepackage{graphicx}
\usepackage{dcolumn}
\usepackage{bm}

\usepackage{microtype}
\usepackage{bm}
\usepackage{appendix}
\usepackage{etoolbox}

\usepackage{epsfig}
\usepackage{graphicx}
\usepackage{amssymb,amsmath,amsbsy,amsgen,amsfonts}    
\usepackage{dcolumn}
\usepackage{amsthm}
\usepackage{mathrsfs}
\usepackage{latexsym}
\usepackage{array}
\usepackage{color}
\usepackage{amstext}
\allowdisplaybreaks[1]
\usepackage{txfonts}
\usepackage{pifont}
\usepackage{braket}
\usepackage{epstopdf} 
\usepackage{color}
\usepackage[dvipsnames]{xcolor}
\usepackage[hidelinks,colorlinks=true, linkcolor=WildStrawberry, urlcolor=WildStrawberry, citecolor=Emerald, linktocpage=true, breaklinks=true, bookmarksopen=true]{hyperref}

\usepackage[USenglish]{babel}
\newcommand{\be}{\begin{equation}}
\newcommand{\ee}{\end{equation}}
\newcommand{\ba}{\begin{array}}
\newcommand{\ea}{\end{array}}
\newcommand{\bqa}{\begin{eqnarray}}
\newcommand{\eqa}{\end{eqnarray}}

\graphicspath{ {Plots/} }
\DeclareSymbolFont{symbols}{OMS}{cmsy}{m}{n}

\begin{document}

\title[]{Experimental plasmonic sensing of malaria using an aluminum metasurface}
\author{A. S. Kiyumbi}
\email{amos.kiyumbi@udsm.ac.tz}
\affiliation{Department of Physics, Mathematics and Informatics, Dar es Salaam University College of Education, PO 
 Box 2329, Dar es Salaam, Tanzania}
\affiliation{Department of Physics, Stellenbosch University, Private Bag X1, Matieland 7602, South Africa}
\author{M. S. Tame}
\affiliation{Department of Physics, Stellenbosch University, Private Bag X1, Matieland 7602, South Africa}
\date{\today}
\begin{abstract}
A wide range of methods currently exist for testing the presence of malaria, each with its own advantages and disadvantages. New technologies are urgently needed to develop more effective diagnosis tools to fight and eradicate malaria. Optical biosensors that employ surface plasmon resonance~(SPR) techniques are a promising category of devices for detecting malaria biomarkers. One such biomarker is plasmodium lactate dehydrogenase~(pLDH), a protein produced during the life cycle of the malaria parasite, which is a metabolic enzyme found in all plasmodium species, including the most widespread falciparum.~This work reports on the design, probing, and experimental performance of an optical biosensor for detecting pLDH based on SPR and extraordinary optical transmission. The biosensor is composed of an aluminum metasurface made from an array of nanoholes. The sensor operates in the visible spectral region and achieves label-free sensing of plasmodium falciparum LDH~(pfLDH) spiked in phosphate-buffered saline. The sensor has a spectral sensitivity of $360$~nm/RIU and an~LOD of 1.3~nM, equivalent to 45.6 ng/mL of pfLDH. This type of optical biosensor may offer a cost-effective and high sensitivity method for active infection diagnosis.
\end{abstract}

\maketitle

\section{\label{Introduction}Introduction}
Malaria is a life-threatening disease and a global health problem claiming more than $500,000$ lives every year~\cite{varo2020update, Phillips2017}. For example, global malaria deaths in 2022 and 2023 were $600,000$ and $597,000$, respectively~\cite{WHO2023,WHO2022}. Malaria cases and deaths are reducing at a very slow rate. The 2023 mortality rate, for instance, was $13.7$ deaths per $100,000$ population at risk, which was far higher than the World Health Organization target rate of 5.5. The target rate by 2030 is an optimistic 1.5 deaths per $100,000$ population at risk ~\cite{WHO2023_state}. The disease mainly affects tropical African countries, with a four times higher mortality rate than the global average. The highest percentage of malaria cases reside in five countries: Nigeria ($26\%$), the Democratic Republic of Congo ($13\%$), Uganda ($5\%$), Ethiopia ($4\%$) and Mozambique ($4\%$)~\cite{WHO2023, WHO2023_state}, where children are a large proportion of those affected. Plasmodium falciparum and vivax are the most widespread and deadly species causing malaria~\cite{varo2020update, Phillips2017, WHO2023}.

Two malaria vaccines, RTS,S/AS01 (Mosquirix)~\cite{laurens2020rts} and
R21/Matrix-M~\cite{aderinto2024perspective}, are currently being rolled out in endemic regions
through routine childhood immunization~\cite{venkatesan2024routine, ogieuhi2024narrative}. The aim is to eliminate and eradicate malaria by 2030~\cite{parums2023current}. However, neither vaccine targets parasitic infections with 100$\%$ efficacy~\cite{datoo2024safety, kisambale2025genetic}. The malaria vaccine RTS,S, for instance, offers only $30-50\%$ protection, which is insufficient to interrupt transmission, and its efficacy diminishes over time~\cite{ogieuhi2024narrative, sibomana2025routine}. Due to the high rate of parasite replication, rapid mutation, complex lifecycle, global funding gap and ineffective diagnosis, large-scale vaccination efforts can be expected to meet significant challenges in endemic regions~\cite{sibomana2025routine}.

Malaria rapid detection tests (mRDTs) are widely used to diagnose malaria in endemic regions and are aiding the above-mentioned vaccine programs. They are immunochromatographic antigen assays specific to histidine-rich protein-2 (HRP-2). The biomarker HRP-2 is the most abundant malaria antigen in infected blood. However, it is produced only by the plasmodium falciparum (pf) species. Therefore, the prevalence of non-pf species and deletion or mutation of the HRP-2 gene are misdiagnosed by mRDTs as false negative~\cite{WHO2019, Orish2018}. Moreover, the sensitivity and specificity of more general \textit{pan}-mRDTs in diagnosing sub-microscopic pf malaria infections or non-pfHRP-2 antigens are low~\cite{Gatton2015}. On the other hand, plasmodium lactate dehydrogenase (pLDH) is an alternative malaria biomarker produced by the active plasmodium parasite~\cite{Jain2014, Ragavan2018}. It is an intracellular protein and final metabolic enzyme of the glycolytic pathway found in all plasmodium species~\cite{Dunn1996, Brown2004, Jain2014}, and therefore its detection avoids the selectivity problem of pfHRP-2.

Furthermore, conventional tools for malaria diagnosis like HRP-2 detecting mRDTs are time-consuming and expensive, requiring trained personnel and centralized laboratory equipment. On the other hand, optical biosensors, especially those operating under surface plasmon resonance (SPR) techniques, are simple to use, offering high sensitivity with a low limit of detection (LOD) and label-free sensing in real time. SPR sensors are commercially available, $\textit{e.g}$. BIAcore\texttrademark, and are used in industry and research laboratories to study chemical and biological interactions, such as antibody-antigen interactions and DNA-drug binding, amongst others~\cite{Homola2008, Homola2006, Homola2003, Soler2019}. Therefore, they have the potential to be used as an alternative tool for detecting submicroscopic malaria infections and non-pfHRP-2 antigens (such as pLDH and aldolase)~\cite{Kumar2024, Chaudhary2021}.

Plasmonic-based biosensing holds great promise for rapid and sensitive
malaria diagnosis -- particularly in regions with limited access to advanced diagnostic tools -- by enabling early detection and monitoring of malaria infection at the point-of-care. These sensors have experimentally shown evidence for detecting low concentrations of pLDH spiked in a buffer solution or whole blood lysate~\cite{Cho2013, Bohdan2020}. The best achieved LOD so far is 18 fM, based on a plasmon-enhanced fluorescence immunosensor~\cite{Minopoli2020}. A recent review on optical and other biosensor assays for the diagnosis of
malaria can be found in Refs.~\cite{Ragavan2018, Krampa2020}. Despite promising results from the studies covered in the review, the majority of plasmonic-based biosensors have been fabricated using gold. A cost-effective alternative to gold for commercialization is aluminum, which is $10,000$ times cheaper and its processing is more mature in semiconductor fabrication plants, making large area fabrication ({\it e.g.} films and nanostructured arrays) more accessible~\cite{Barrios2015}. 

Aluminum is both easy to fabricate and functionalize, with material properties that enable plasmon resonances across a broad optical band, in the UV-visible range ($200-700$ nm), due to its negative real permittivity ($-5$ to $-60$) and positive imaginary permittivity (1 to 30)~\cite{West2020, Davy2015}. Compared to gold (which spans the visible-NIR range and has broader SP resonances with a Q-factor $\sim$10--20) and silver (spans the visible-NIR range, with the sharpest resonances and Q$\sim$50), aluminum with a suitable patterning of nanostructures (such as in a photonic crystal fibre~\cite{Ramola25,Ramola25b,Ramola25c,Ramola25d} or nanohole array~\cite{Davy2015}), has moderate sharp resonances (Q$\sim$20--30) associated with propagating surface plasmons and localised plasmons. It also has a superior UV extension that can be an advantage when using analytes with UV absorption/fluorescence. 

Furthermore, passivated aluminum (e.g., SiO$_2$ coated) is non-toxic like gold, while silver is antibacterial but cytotoxic and has rapid degradation due to bulk oxidation. Aluminum shows no evidence of bulk oxidation -- an oxide layer naturally forms on the aluminum surface during air exposure (if not passivated) and remains consistently stable for many days, effectively serving as a natural passivation layer that inhibits further oxidation.~\cite{langhammer2008localized, Davy2015, zhang2018long, martin2014Al_fabrication}. Further details on the fundamental plasmonic properties of aluminum and how they compare to other plasmonic materials like gold and silver can be found in recent review articles~\cite{Lambert20,Li23,Shukla25}. \\

In this work, we report on the design and experimental probing of a plasmonic-based malaria sensor using an aluminum metasurface. The sensor is an array of nanoholes and operates in the visible spectral region. It provides label-free sensing of pLDH, which we test by spiking pLDH into phosphate-buffered saline (PBS). We measure a sensor sensitivity of $360$~nm/RIU (refractive index units), an LOD of $1.3$~nM and an equilibrium binding affinity of $71$~nM for the pfLDH antigen interaction with its IgG antibody. The LOD of $1.3$~nM achieved using aluminum nanoholes is better than the LOD of $ 23.5$~nM previously reported using gold nanoholes~\cite{Bohdan2020}. Our results may therefore open up new possibilities for designing cost-effective and highly sensitive malaria sensors based on plasmonics. As the antibody we use also binds to pLDH from other plasmodium species, the detection of pfLDH and other species pLDH in real blood samples in the form of whole blood lysate would be the next step for the sensor's development.

The paper is structured as follows: in Section II, we provide the physics of our sensor. In Section III, we describe the optical setup, the experimental methods used, and the functionalization protocol followed to develop the plasmonic-based malaria sensor. In Section IV, we provide our experimental results and compare them with the calculated results based on a finite element method (FEM) using the software COMSOL. In Section V, we provide a summary and an outlook on future~work.

\section{\label{Biosensor model}Biosensor model}
Plasmonic biosensing utilizes confined surface plasmons (SPs) to detect the presence of analytes, $\textit{e.g}$. proteins and enzymes, in a blood sample or a buffer solution. The SPR sensing protocol offers a highly sensitive and rapid method for early infection detection by monitoring changes in light interaction when a specific biomarker binds to a specially designed sensor surface coated with receptors, $\textit{e.g}$. antibodies, aptamers, etc. Plasmonic metals such as gold, silver, copper, and aluminum support the excitation of SPs in the visible optical regime~\cite{Maier2007}, overlapping well with conventional light sources. Here, a photon of appropriate energy and momentum couples to an SP, which is a quasi-particle consisting of a conduction electron density with an electromagnetic evanescent wave at a metal-dielectric interface~\cite{Maier2007}. SPs are excited only if the real part of the metal permittivity has an opposite sign to that of the dielectric medium at the interface. The propagation constant of SPs at the interface is then given~by~\cite{Maier2007}
\be
k_{sp}= k_0 \sqrt{\frac{\varepsilon_m\varepsilon_d}{\varepsilon_m + \varepsilon_d}},
\label{eq:SP}
\ee
where $k_0 = \frac{2\pi}{\lambda}$ is the wavenumber (propagation constant) of the incident light in free-space, $\lambda$ is the free-space wavelength, $\varepsilon_m$ is the permittivity of the metal and $\varepsilon_d = n_d^2$ is the permittivity of the dielectric on top of the metal, with $n_d$ as the refractive index of the dielectric layer. To satisfy mode-matching conditions between the incident light ($k_0$) and SPs ($k_{sp}$), a prism is usually~used~\cite{Maier2007}. 

\begin{figure*}[t]
\centering
\includegraphics[width=18cm]{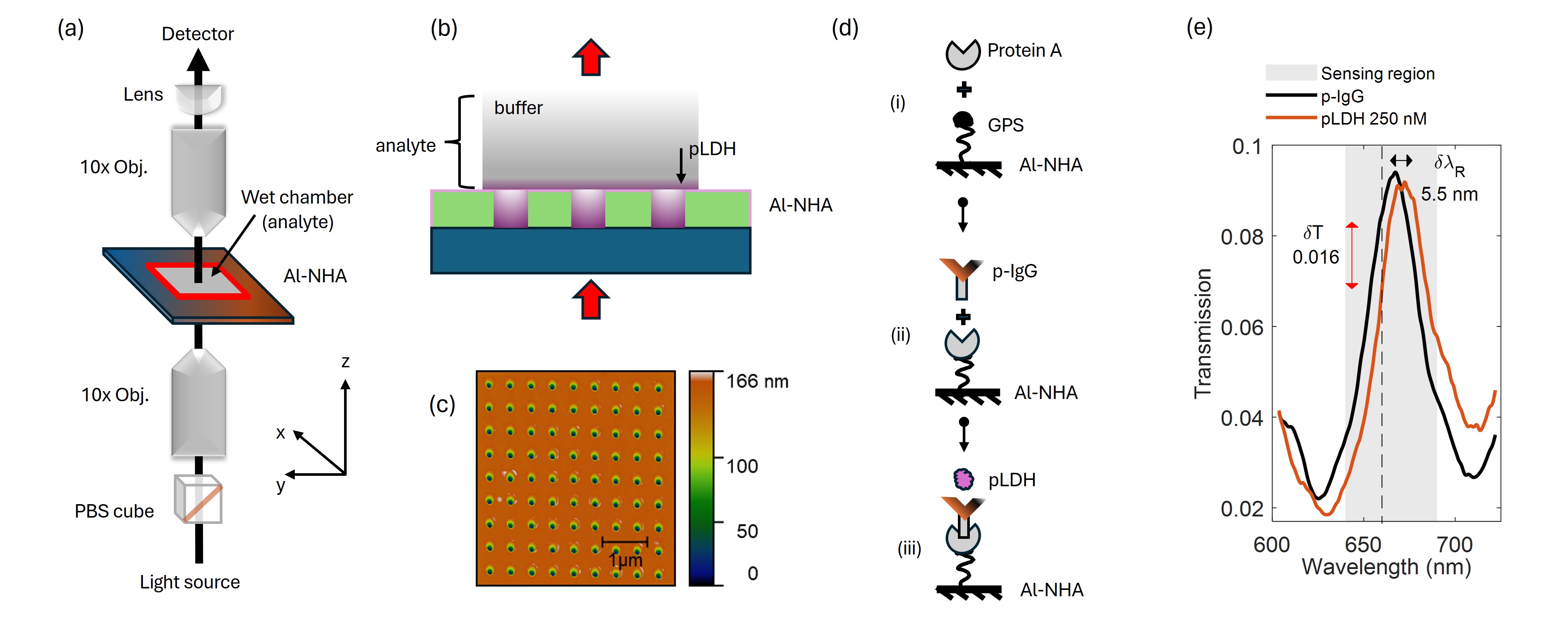}\caption{ Experimental setup, surface chemistry protocol, and measurement used for plasmonic biosensing of the malaria biomarker pfLDH. (a) The optical setup used to probe the sensor. It consists of a collimated broadband light source, a polarizing beamsplitter cube,  two $10\times$~microscope objectives, a plano-convex lens, and a detector -- spectrometer connected to a computer. A nanohole array (NHA) sensor is placed between the two objectives. One objective is used to focus light onto the sensor and the other is used to collect the light transmitted through the sensor. An arrow (solid black line) shows the direction of light propagation. (b) The wet chamber enclosing the sensor. The chamber is filled with analyte solution (buffer + pfLDH) before the sensor is probed. pfLDHs are concentrated closer to the sensor surface as they bind to specific antibodies immobilized on the surface. (c) The atomic force microscope (AFM) image of the Al-NHA. The sensor is made of circular nanoholes in a square lattice, with a period $a_0$ of $450$~nm. The sensor thickness is approximately $166$~nm. (d) The surface functionalization of the sensor. (i) The SiO$_2$ layer is hydroxylated and coated with (3-Glycidoxypropyl)trimethoxy silane (3-GPS), then protein A is anchored to the sensor~through amine bonding. The 3-GPS forms a strong Si-O-Si covalent bond with the hydroxyl group on the SiO$_2$ surface. The epoxide group of 3-GPS reacts with amine nucleophiles of protein A, forming a strong covalent bond~\cite{silberzan1991silanation}. (ii) Antibodies (p-IgG) were then immobilized on the sensor through protein A~\cite{makaraviciute2013site}. (iii) After functionalization, biomarker pLDH binds to the antibody p-IgG. (e) When pfLDH solution is injected into the~wet chamber, the pfLDH molecules bind to p-IgG immobilized on the sensor surface causing a shift of the EOT resonance peak. A~$\delta\lambda_R=5.5$~nm spectral shift from the baseline peak (p-IgG peak) and $\delta T=0.016$ transmission drop at $\lambda=660$~nm (dashed line) is achieved when $250$~nM of~pfLDH is added.}
\label{fig1} 
\end{figure*}

When a plasmonic metal film is perforated with sub-wavelength holes forming a nanohole array (NHA), at a particular wavelength $\lambda_R$, the incident light is extraordinarily transmitted~\cite{Ebbesen1998}. SPs are responsible for this extraordinary optical transmission (EOT)~\cite{Beijnum2012, Popov2000, Moreno2001, Liu2008, Liu2010, lalanne2009}. The transmission is contrary to the classical diffraction law of Bethe for a sub-$\lambda$ hole in a perfect electric conductor~\cite{bethe1994}, where the transmission efficiency $T$ decreases dramatically according to $T \sim (r/\lambda)^4$. For a plasmonic metal, the periodic arrangement of nanoholes provides the necessary momentum to overcome the mismatch between the wavenumber $k_0$ of the incident light and the SP wavenumber $k_{sp}$ given in Eq.~\eqref{eq:SP}~\cite{Maier2007, Popov2000}. The NHA introduces a reciprocal lattice vector that provides the extra momentum $\vec{G}_{xy}$ required for phase matching. The phase matching condition for SP excitation in NHAs is given by 
$\vec{k}_{sp}= \vec{k}_{||} + \vec{G}_{xy}$~\cite{Maier2007, Popov2000},
where $k_{||} = k_0n_d\sin\theta$ is the in-plane ($x-y$ plane) wavevector component of the incident light, and $\theta$ is the angle of incidence from the $z-$axis. $\vec{G}_{xy}$ is the reciprocal lattice vector of the NHA and is determined by the lattice constant $a_0$. For a square lattice, $\vec{G}_{xy} = m(2\pi/a_0)\hat{x} + n(2\pi/a_0)\hat{y}$, where $m$~and~$n$ are integers representing the diffraction orders in the $x$ and $y$ directions~\cite{Maier2007, Popov2000}. At normal incidence, $\theta = 0$, the incident wave can satisfy the condition $k_{sp} = G_{xy}$ to excite an SP, and an EOT peak occurs at wavelength $\lambda_R$ given by~\cite{Maier2007, Ghaemi1998}
\be
\lambda_{R} = a_0(m^2 + n^2)^{-1/2}\sqrt{\frac{\varepsilon_m\varepsilon_d}{\varepsilon_m + \varepsilon_d}}.
\label{eq:PMC3}
\ee
The ($m,n$) $=$ ($1,0$) SP mode is excited by incident light polarized along the $x$ direction, similarly the ($m,n$) $=$ ($0,1$) SP mode is excited when light is polarized along the $y$ direction. SPs excited by the NHA become localized at the corner edges of a nanohole, forming a discrete resonant mode -- a narrowband feature. The mode constructively and destructively interferes with a direct transmission mode comprised of light evanescently tunneling through the holes -- a broad continuum mode (non-resonant background). The interference forms a Fano-like resonance characterized by an asymmetric line shape in the transmission spectrum~\cite{Popov2000, Genet2003}. EOT peaks characterized by $\lambda_R$ are very sensitive to small changes in the refractive index of the dielectric close to the metal surface, as defined in Eq.~\eqref{eq:PMC3}. They therefore provide a means to use a NHA as an optical biosensor through monitoring of its spectral response. 

To a first approximation the coupling efficiency of incident light to SPs on the NHA sensor is determined by the grating momentum matching condition $k_{sp} = G_{xy}$ and how close it is satisfied. When loss and variation in nanohole geometry are included, the condition becomes broadened, with optimal coupling efficiency depending on how close to the plasmonic resonance the wavelength of the incoming light is. The grating momentum matching condition approach is a simple model however. A more rigorous approach is FEM simulation, resulting in a comprehensive spectral profile for the transmission, with peaks roughly at the locations predicted by the simple grating condition approach. A better coupling efficiency makes the transmission peak higher and improves the sensitivity, as the gradient can be larger. This then influences the sensor's LOD (at a fixed amount of noise), as a larger gradient gives a smaller LOD. Further details of this behavior are discussed later.

\section{Experimental scheme}
Aluminum has its strongest plasmonic activity in the ultraviolet and lower visible regions, typically offering an SPR peak between $300-450$~nm~\cite{West2020, Davy2015}. However, it can still support SPR within the entire visible range in structured forms like a NHA ~\cite{Barrios2015}. In this work, an aluminum NHA (Al-NHA) is designed to achieve SPR in the $600-700$~nm optical range by carefully tuning the hole diameter, periodicity, and array geometry. This tuning allows the SP resonance to shift into an optimal sensing region that can be probed with conventional light sources. However, when using aluminum as a plasmonic material, managing oxidation and corrosion is critical because it can alter the position of the SPR resonance and dampen the plasmonic response over time~\cite{Barrios2015}. A thin protective silica SiO$_2$ layer was applied to our NHA, keeping the resonance stable around the targeted $600-700$~nm range. While gold (Au) remains the standard metal for stable plasmonic sensing in the optical region, aluminum is a cost-effective alternative with careful design and oxidation control, enabling accurate analyte detection and quantification~\cite{martin2014Al_fabrication, Victor2014, Lee2017}.

Since a NHA supports EOT-SPR peaks that shift depending on the local dielectric refractive index, a metasurface made of a NHA can use these peaks for sensing changes in the NHA environment due to antibody-antigen interactions, as discussed in Section II. This sensing technique essentially monitors the shift of transmission peaks or intensity associated with bound analytes in the close vicinity of the localized SP fields. The approach is different from the conventional Kretschmann SPR sensor~\cite{shukla2022surface}. EOT sensors exclude the need for a prism and offer comparable high sensitivity with more flexibility, while allowing for better integration and multiplexing~\cite{Lee_2017, Homola1999b}. 

\subsection{Sensor and experimental setup}
In Fig.~\ref{fig1}(a), we show the experimental setup used for plasmonic sensing of pLDH for the most widespread falciparum species of the malaria parasite, $i.e.$ pfLDH. It consists of a broadband white light source (MBB1F1 from Thorlabs), polarizing beamsplitter (WPBS254-VIS from Thorlabs), achromat objectives (10x, 0.25 NA, 10.6 mm WD), malaria sensor (Al-NHA), plano-convex lens ($f =50$ mm) and a detector system. The collimated light from the white light source passes through the polarizing beamsplitter before it is focused on the sensor. The beamsplitter reflects the s-polarized component of the incident light and transmits the p-polarized component to the sensor. The p-polarized light is focused on the sensor using a $10\times$ objective. The transmitted light is collected by another $10\times$ objective placed at an equal distance ($10.6$ mm) from the sensor. A plano-convex lens receives the transmitted light beam and focuses it on a fiber coupler. The coupler is connected to a compact CCD spectrometer (CCS200 from Thorlabs), which is connected to a computer for spectral analysis.

Our Al-NHA sensor is operated under static conditions, where there is no fluidic motion of the buffer and the binding dynamics have reached equilibrium, see Fig.~\ref{fig1}(b). Different concentrations of pfLDH (REC 31737 from the Native Antigen Company) were prepared using PBS (pH 7.4, liquid, sterile-filtered from Merck) and injected into the wet chamber. The wet chamber is a silicone cavity with dimensions $20\times 20\times 1$ mm, which encloses the sensing area. pfLDH molecules diffuse from the bulk solution and bind to p-IgG (MAB 12350 from the Native Antigen Company) immobilized on the sensor surface, forming a layer of protein on top of the antibodies. The p-IgG we use has the potential to bind to pLDH produced by all plasmodium species, although in this work, as a first study, we focus on binding and detection of pfLDH. The choice of antibody was motivated by its availability, verified specificity for pfLDH, and relevance for benchmarking against prior work (Refs.~\cite{Cho2013,Bohdan2020}). 

The effective refractive index, $na_{\text{eff}}$, of the protein layer depends on the bulk concentration of pfLDH and the binding affinity of the interaction. When the sensor is probed, it excites SPs at the alumina interfaces. The localized SP fields enclose the adsorbed protein layer, and any small change in $na_{\text{eff}}$ of the thin layer due to a change in pfLDH concentration of the bulk solution results in a spectral shift of the EOT peak~\cite{Cho2013}. 

\subsection{Al-NHA fabrication}
The Al-NHA was fabricated by Moxtek Inc.\ to specification using UV nanoimprint lithography (UV-NIL) on a glass substrate. The UV-NIL followed by Moxtek is described in Refs.~\cite{George14,George15}, and is a standard and well-documented method in the field of nanostructure fabrication~\cite{martin2014Al_fabrication}. A proprietary passivation layer (SiO$_2$) is applied to the NHA to prevent oxidation~\cite{west2010searching, Davy2015}. The SiO$_2$ layer effectively prevents aluminum oxidation, which would otherwise form a native Al$_2$O$_3$ layer of $2.5-3$ nm thickness upon air exposure~\cite{langhammer2008localized, zhang2018long}. Such oxidation reduces the plasmonic scattering efficiency, broadens and red-shifts the resonance peak, and decreases refractive index sensitivity, leading to a degraded sensor performance with reduced spectral sensitivity and higher LOD~\cite{zhang2018long, Victor2014, Davy2015}. By preventing oxidation, the original plasmonic properties are maintained for optimal biosensor operation~\cite{Davy2015, Barrios2015}. 

In Fig.~\ref{fig1}~(c), we show the topography of our Al-NHA sensor using an atomic force microscope (AFM). The sensor is made up of circular nanoholes in a square lattice. The Al surface is coated with a $5$~nm conformal SiO$_2$ layer, added on top of the nanoholes, and a $10$~nm TiO$_2$ adhesion layer between the aluminum and the glass substrate. We used the AFM topography and a scanning electron microscope (SEM) image (see Appendix~A for details) of the Al-NHA to confirm the geometric parameters. The AFM scan confirms a period of $450$~nm and a NHA thickness of $166$~nm. The SEM scan confirms that the size of holes within the array has, on average, a hole diameter $d = 242$~nm. These geometric parameters match closely with those requested ($a_0 = 450$~nm, $t = 150$~nm, $d = 250$~nm), which were chosen to give an EOT peak that overlapped with the spectrum of our~white~light~source (600-700~nm). 

Our SEM and AFM analysis gives an indication of the consistency of the nanoholes in terms of variations. Furthermore, we measured several different NHA’s with the same specification and found similar results for the dimensions. The UV fabrication process used is known to give a high level of repeatability and high yield, as described in more detail in Refs.~\cite{Fruncillo21,Stokes23}, which our measurements confirm. This is an important aspect for eventual large-scale production and commercialisation of the sensor.

In Fig.~\ref{fig1}~(d), we show how the Al-NHA is turned into a malaria biosensor by functionalizing the SiO$_2$-coating surface with antibodies that detect pfLDH. The process starts with silanization~\cite{Hermanson2013chap13, Hermanson2013chap15, silberzan1991silanation}. The SiO$_2$ passivated Al-NHA was sonicated in isopropanol for 15 minutes and cleaned in toluene, air-dried and hydroxylated through O$_2$ plasma treatment. The NHA was then coated with 3-GPS, a silane coupling agent. The 3-GPS-coated Al-NHA was then heated in the oven at $130^{\circ}$ C and sonicated in toluene for 15 minutes. After washing to remove toluene traces, the NHA was conjugated to 1~mg/mL of protein A in PBS at room temperature overnight, then incubated in $50$ $\mu$g/mL p-IgG solution in PBS for four hours at room temperature. Protein A binds the Fc region of the IgG antibodies~\cite{Surolia1982, Galanti2016}, orienting the Fab region for optimal interaction with a target analyte ($\textit{e.g}$. pfLDH)~\cite{makaraviciute2013site}. More details on the sensor functionalization are given in Appendix~B.

\subsection{Measurement analysis background}
The analyte used in this study is a recombinant plasmodium falciparum LDH protein (pfLDH, gene accession: DQ198262.1) from the Native Antigen Company that was purified from E. coli ($\ge 95\%$ pure) and presented in a liquid form~\cite{kleiner2018secretion}. The initial stock solution had a concentration of $1.31$ mg/mL. To prepare a $250$~nM solution of pfLDH from the $1.31$ mg/mL stock solution (using a MW of $35$ kDa~\cite{Jain2014}), we diluted approximately $6.7$ $\mu$L of the stock with $993.3$ $\mu$L of PBS to make a final volume of $1$ mL. Only $50$ $\mu$L of this solution was used in each measurement. Other pfLDH solutions of lower concentration were obtained by serial dilution of the $250$~nM solution. For sensing, the pfLDH solution was injected into the wet chamber using a $10 - 100$ $\mu$L pipette for precise and accurate liquid handling. 
\begin{figure}[t!]
\centering
\includegraphics[width=8.2cm]{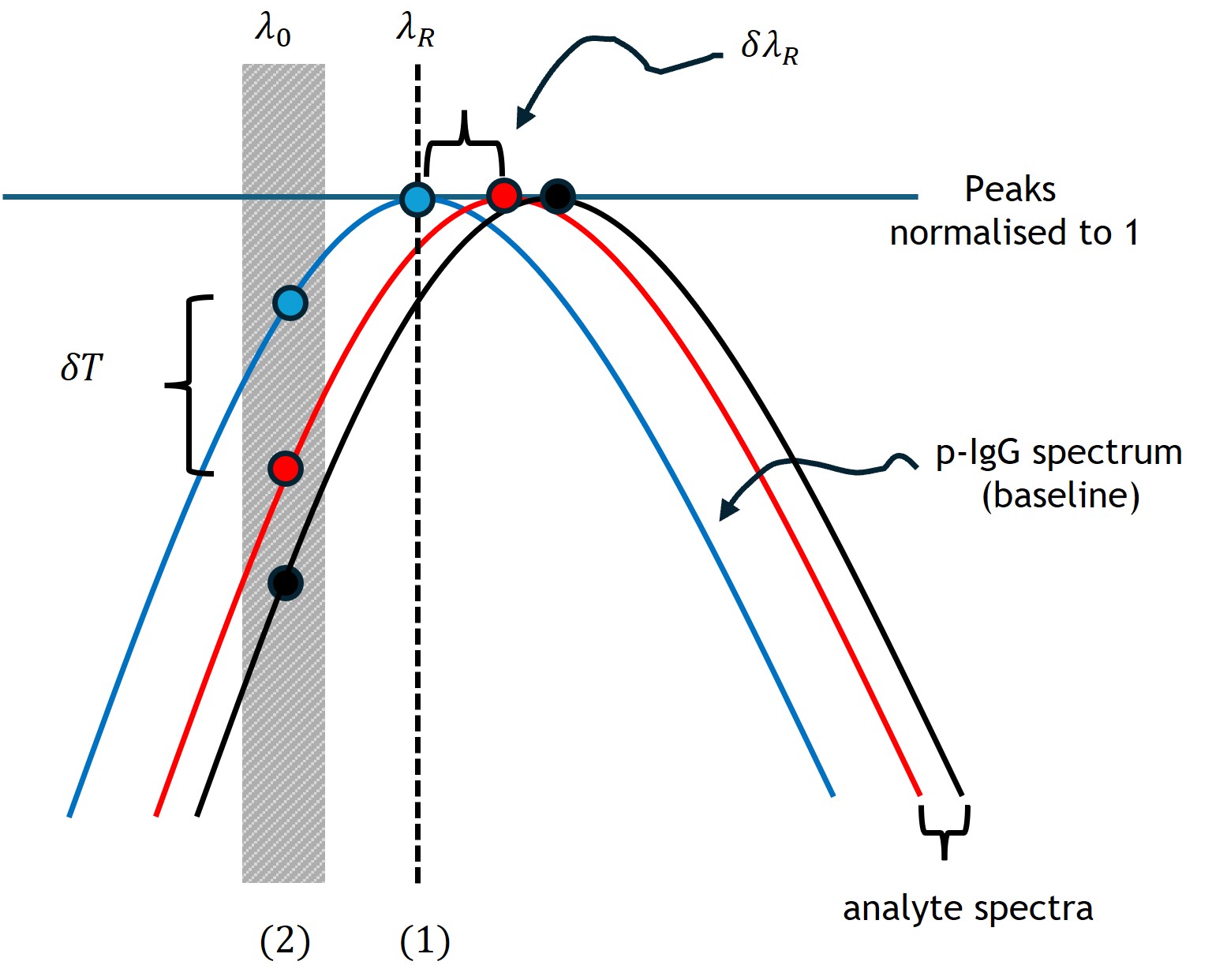}
\caption{A schematic showing the spectral sensing and intensity sensing modality used in this study. $\delta\lambda_R$ is a spectral shift of the EOT peak measured from the baseline peak -- the p-IgG blank spectrum. $\delta T$ is the drop of transmission measured from the p-IgG spectrum~at~a~fixed wavelength $\lambda_0$. (1) and (2) represent intensity sensing evaluated~at~the p-IgG baseline peak at $\lambda_R$ and off the baseline peak at $\lambda_0$, respectively.}
\label{fig3} 
\end{figure}

To give a uniform spectrum across the wavelength range of interest, $450-900$~nm, the measured spectra are normalized to the light source spectrum -- light through the optical components of the setup in the absence of the sensor, which we call the reference spectrum, $T_R$. The normalized transmission, $T$, of the sensor response is then calculated as 
\be
T = \frac{T_s}{T_R}\left(\frac{t_R}{t_s}\right),
\label{eq:T}
\ee
where $T_s$ is the pfLDH measured spectrum (not normalized), $t_R$ is the integration time of the reference spectrum ($t_R = 37.4$ ms) and $t_s$ is the integration time of the pfLDH measured spectrum ($t_s = 378$ ms). Background spectra were collected and subtracted from the measured spectrum. Each spectrum is calculated as an average of ten measured spectra. It is important to note that the calculated transmission spectrum, $T$, contains noise. The noise level depends on the power of the source of light and the integration time of the detector, as well as technical noise, such as from the spectrometer. Therefore, a moving average algorithm in Matlab is used to smooth the raw data, and a spline fitting function is applied to help locate the EOT spectral peak position. In Matlab, the moving mean algorithm is described as $m = \text{movmean}(T,k)$, where $m$ is an array of local $k-$point mean values, where each mean is calculated over a sliding window of length $k$ across neighboring elements of $T$~\cite{Matlab2022}.

In Fig.~\ref{fig1}~(e), we show one of our pfLDH sensing results, the case of $250$~nM pfLDH solution injected into the wet chamber's cavity region. It can be clearly seen that the sensor is able to distinguish the pfLDH solution from the one without pfLDH. A shift in the spectral resonance $\delta \lambda_R$ of $5.5$~nm and a transmission drop $\delta T$ of $0.016$ (evaluated at $\lambda = 660 $~nm) are observed within~$5-10$~minutes~of~the~injection~time.

\subsection{Plasmonic sensing scheme}
In Fig.~\ref{fig3}, we present the two sensing modalities analyzed in this study: spectral sensing and intensity sensing. While spectral sensing tracks shifts in the EOT resonance peak in the spectrum $\delta\lambda_R$, intensity sensing monitors changes in the transmittance $\delta T$ at a fixed wavelength. Changes in the local refractive index due to pfLDH binding to IgG near the sensor surface modify the coupling efficiency of light to the plasmonic modes, thus altering the intensity of transmitted light and shifting the EOT peak position. For convenience, the sensor spectra are normalized to $1$ using the formula 
\be
T_{\text{norm}} = T/T_{\text{max}}, 
\label{eq:T2}
\ee
where $T_{\text{max}}$ is the maximum transmittance at the EOT peak achieved by the sensor for a particular concentration. While spectral sensing simply measures the shift of the peak $\delta\lambda_R$, for intensity sensing, we monitored the transmission at two places: (1) at the fixed wavelength $\lambda_R = 667$~nm at the peak, and (2) at fixed wavelengths $\lambda_0 = 650$~nm, $655$~nm, and $660$~nm off the peak, shown as the gray region in Fig.~\ref{fig3}. The transmittance value of the p-IgG spectrum at these particular wavelengths is used as a baseline. The off-peak resonance wavelengths are wavelengths on the steep left-hand side of the chosen resonance EOT peak, where the sensitivity is the highest and thus the LOD is expected to be the lowest. A few different values were chosen to see the variation in sensing performance, with 660 nm expected to be optimal (the highest gradient) from simulations, but it was important to check the variability in performance around this wavelength experimentally. It turns out that just below 660 nm, at 655 nm, the sensitivity is optimal, due to non-uniformity of the peak shape as it shifts in the experiment compared to the behaviour expected from simulations. 

The sensor sensitivity $S_R$ (where $R$ represents the sensor response, $R=\lambda_R \; \text{or} \:T$), and LOD of the two sensing schemes are calculated as~\cite{Maier2007, Homola2008, Spackova2016}
\be
S_R = \frac{\delta R}{\delta n_{\text{eff}}},
\label{eq:Sens}
\ee
and~\cite{Armbruster2008, Homola2008, Spackova2016}
\be
\text{LOD} = \frac{\sigma_R}{S_R},
\label{eq:LOD}
\ee
where $\sigma_R=3\Delta R$ and $\Delta R$ is the noise associated with the measurement of $R$. The sensor sensitivity refers to the sensor's ability to detect and respond to changes in the quantity it is designed to measure. In the above, $\delta R$ is the change in sensor response (wavelength or transmission) due to a change in the effective refractive index, $n_{\text{eff}}$, in the region above the sensor. On the other hand, the LOD is the smallest change in effective refractive index, $\Delta n_{\text{eff}}$, that can be reliably detected by the sensor. $\Delta R$ is the standard deviation of $R$ for a blank ($0\%$ pfLDH) analyte solution. It is a measure of the noise level in the sensor system when no analyte is present. The factor of 3 is commonly used in most sensing schemes to ensure that the sensor response is distinguishable from noise with $99\%$ confidence~\cite{Armbruster2008, Spackova2016}. Note that $S_R$ and the LOD can also be defined in terms of changes in concentration. Assuming a linear sensor response for low concentration, $L_0$, of protein pfLDH in the buffer, we have the LOD as $\Delta L_0 = {\sigma_R}/{S_R}$. Here, $\Delta L_0$ is the lowest protein concentration that can be detected in the presence of a signal noise $\sigma_R$ and $S_R = {\delta R}/{\delta L_0}$ is the sensor sensitivity, defined as the sensor response for a given change in the bulk concentration. A high $S_R$ and a low $\sigma_R$ means a better LOD and therefore~the~sensor~can detect smaller changes in concentration.

\section{Results}
\subsection{Transmission measurement results}
\begin{figure}[t!]
\centering
\includegraphics[width=9cm]{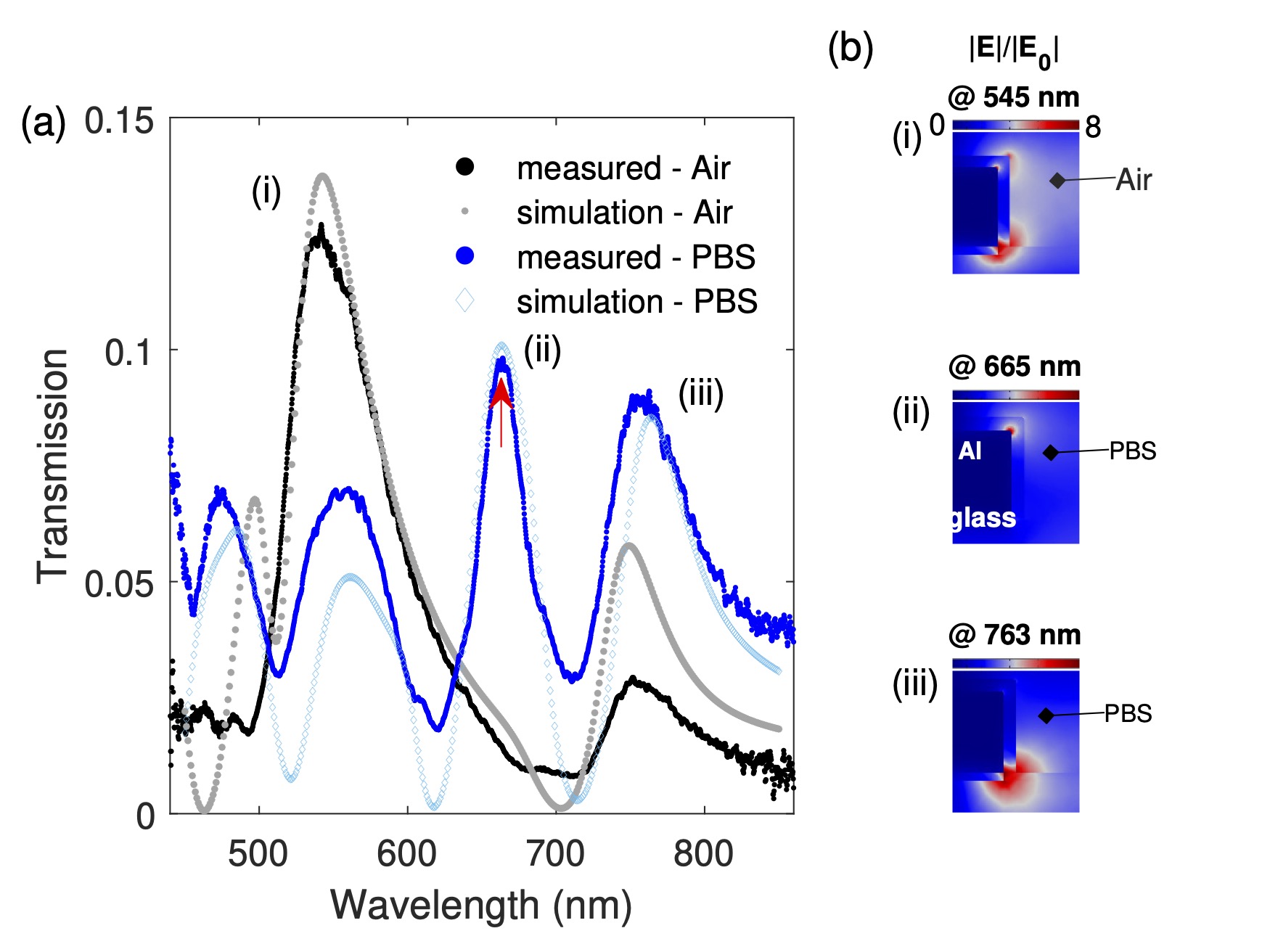}
\caption{Transmission spectra of a non-functionalized sensor and the near-field distribution inside one of the nanoholes. (a) Measured spectra (in air and PBS) compared to finite element method (FEM) simulations. Peak (i) shows the spectral resonance of the sensor in air ($\lambda_R = 545$~nm for both experiment and simulation), and (ii) shows the spectral resonance of the sensor in PBS ($\lambda_R = 664$~nm for experiment and $665$~nm for simulation). Peak (iii) is the spectral peak associated with the aluminum-interband transition ($\lambda_R = 760$~nm for experiment and $763$~nm for simulation). The red arrow points at the sensing peak used in this study, as it was found to be the most sensitive due to a significant presence of the electric field $E$ in the analyte region. (b) A cross-section of a half nanohole cell shows the electric field distribution and enhancement $|E|/|E_0|$ at the resonance~peaks indicated~in panel~(a). $E_0$ is the electric field amplitude of the incident plane wave.}
\label{fig4} 
\end{figure}

In Fig.~\ref{fig4}~(a), we show the measured spectra of the non-functionalized Al-NHA sensor when in air and when the wet chamber is filled with PBS. The spectra are compared with FEM simulated spectra using COMSOL (see Appendix~B for FEM simulation details). For modeling, the refractive index of PBS was taken as $n_{\text{eff}} =1.335$ and that of glass as $n_s = 1.52$. It can be seen that the measured spectra are clearly in line with the simulations. The experiment spectra profiles, number of peaks, and peak positions are correctly reproduced. For example, for the PBS spectrum, at the peak labeled (ii), there is a spectral difference of only 1~nm between the experiment ($\lambda = 664$~nm) and simulation ($\lambda = 665$~nm). Moreover, the close match between the experiment and simulation confirms that the simulation parameters match well with the fabricated Al-NHA parameters and that the simulation can be used to predict with high confidence the sensor's optical behavior. We tested several fabricated samples and found consistent results in terms of the peak locations ($<1$~nm), width ($<1$~nm) and height ($<1-2$~\%).

In Fig.~\ref{fig4}~(b), we show a cross-sectional view of the optical field distributions within the Al-NHA. A half of a unit cell is displayed. The electric field enhancement $|E|/|E_0|$ of the three simulated peak wavelengths: peak (i) $\lambda_R = 545$~nm, peak (ii) $\lambda_R = 665$~nm, and peak~(iii)~$\lambda = 763$~nm, are shown. $E_0$ is the electric field amplitude of the incident plane wave. A detailed study of peak~(i) based on the field distribution confirms that it is made of two plasmonic degenerate modes, one due to SP excitation at the glass/aluminum interface localized at the bottom corner and the other due to SP~excitation at the analyte/aluminum interface localized at the top edge corner. In the spectral range~of~$\lambda = 500 - 600$~nm, the two modes have the same or nearly the same energy (frequency), resulting in a~single~peak~in~the~transmission~spectrum~in~air~($n_{\text{eff}} = 1.00$).\\

\begin{figure}[t]
\centering
\includegraphics[width=9cm]{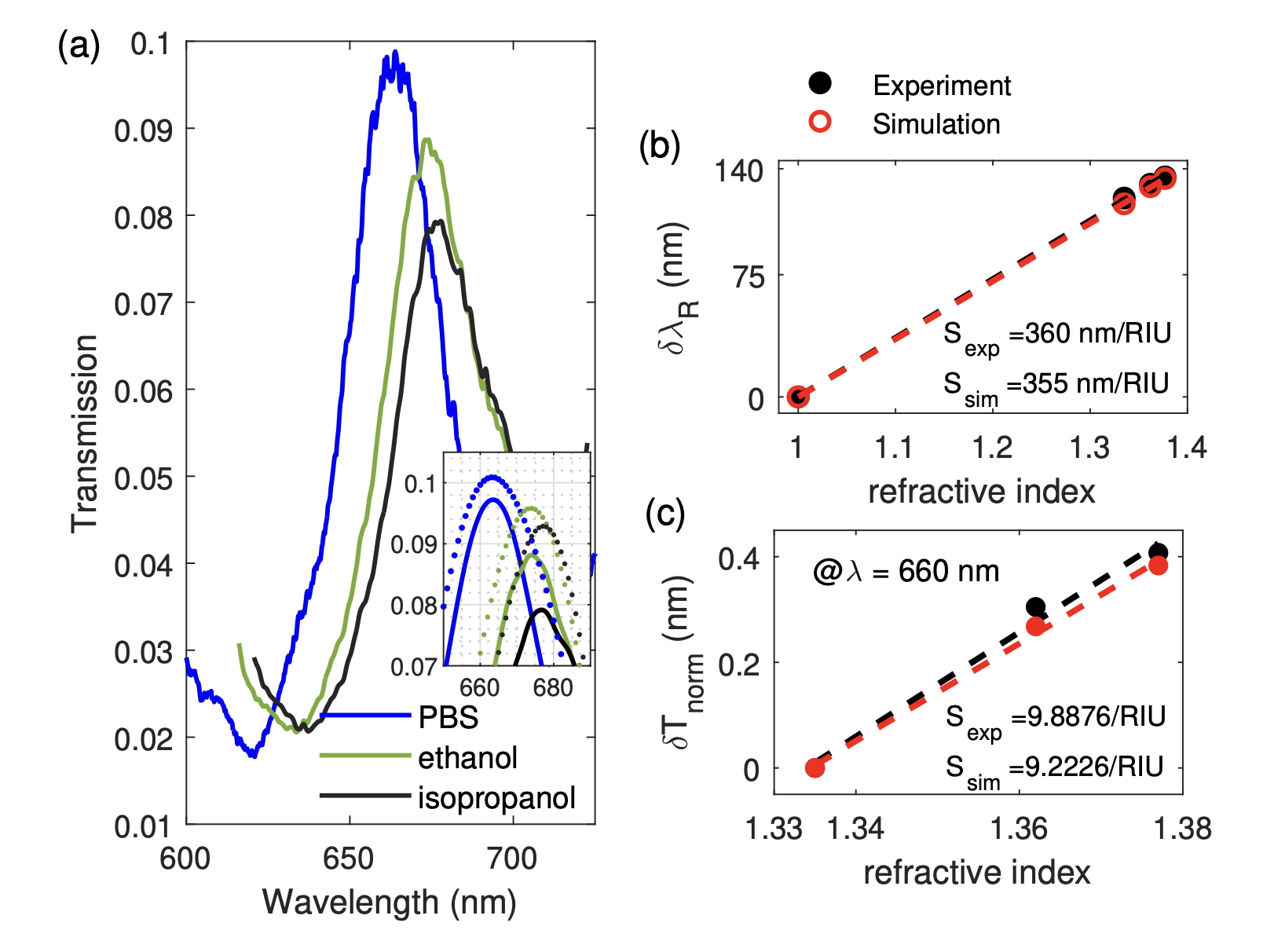}
\caption{Sensor calibration using bulk organic solvents. (a) Measured transmission of the non-functionalized sensor in PBS ($n_{\text{eff}} = 1.335$), ethanol ($n_{\text{eff}} = 1.36$2), and isopropanol ($n_{\text{eff}} = 1.377$). The inset shows a zoomed-in spline fitting of the measured data (solid lines) and the corresponding FEM simulated spectra (dotted lines). The bulk sensitivities are calculated for panel (b) with spectral sensing at resonance (including $n_{\text{eff}}=1.00$ for air) and panel (c) with intensity sensing off-resonance at $\lambda_0 = 660 $~nm (the air resonance peak is too far away to be used). A sensitivity of $360 \;(355)$~nm/RIU and $9.8876 \;(9.2226)$ a.u./RIU are obtained based on the experimental data~(FEM simulation) for spectral and intensity sensing, respectively.}
\label{fig5} 
\end{figure}

When $n_{\text{eff}}$ is increased from $1.00$ (air) to $1.335$ (PBS) in steps of $0.05$ RIU, peak (i) decomposes into two modes: as  $n_{\text{eff}}$ increases, the coupling strength between the evanescent NHA modes, SP modes, and directly transmitted modes, is enhanced and the hybrid mode starts to split. One mode remains at the original position, $\lambda \sim 545$~nm, and the other redshifts to a higher wavelength. When $n_{\text{eff}}$ reaches $1.335$ (PBS), the two degenerate modes become completely separate and distinct, one at $\lambda = 555$~nm -- close to peak (i), and another at $\lambda = 665$~nm -- away from peak (i). The former is related to SP modes at the glass/aluminum interface, and the latter,  at $\lambda = 665$~nm, is related to SP modes at the analyte/aluminum interface, and~is~presented as peak (ii) in~Fig.~\ref{fig4}~(a) and (b).

As a result of the above observation, peak (ii) is identified as the most sensitive to refractive index changes and is the best EOT peak to monitor the NHA sensor behavior when various pfLDH concentrations are added to a suitable buffer like PBS. The fields are more concentrated on the analyte side, and therefore a small perturbation in the analyte layer refractive index, $na_{\text{eff}}$, can be more easily tracked using peak (ii) than any other transmission peak~\cite{Wu2012}. Peak (iii), at $\lambda = 763$~nm, is related to the intrinsic property of the aluminum metal. Since most of its near field is concentrated on the glass side, the peak responds less to changes in $na_{\text{eff}}$. This peak is not as optimal for sensing applications and represents the interband transition of surface electrons in aluminum~\cite{Barrios2015, Davy2015}. Interband transitions in aluminum occur at photon energies greater than $1.5-1.6$ eV (corresponding to a wavelength $\sim 800$~nm). They can be characterized by a reduced EOT peak intensity, broader spectral feature and a sharp increase in absorption.

To study the impact of the 5~nm SiO$_2$ passivation layer in the simulations, we modelled the system with and without the layer. We found that it resulted in a $<1$~nm red shift of the resonance peaks and negligible change in peak height. Thus, the layer protecting the aluminum from oxidation has only a small impact on the sensor's behavior. As the SiO$_2$ layer prevents oxidation we expect little oxidation takes place in the experiment and so we can avoid effects that are usually associated with the build-up of an oxide layer, such as the broadening of the resonances, the lowering of the peak heights and their redshift due to the loss of support of plasmonic excitations. These effects for aluminum are discussed in more detail in Refs.~\cite{Barrios2015,Li14}. The overall impact of oxidation effects would be a drop in the sensitivity and a subsequent increase in the LOD. 

\subsection{Sensor calibration: bulk sensing}
\begin{figure}[t]
\centering
\includegraphics[width=9cm]{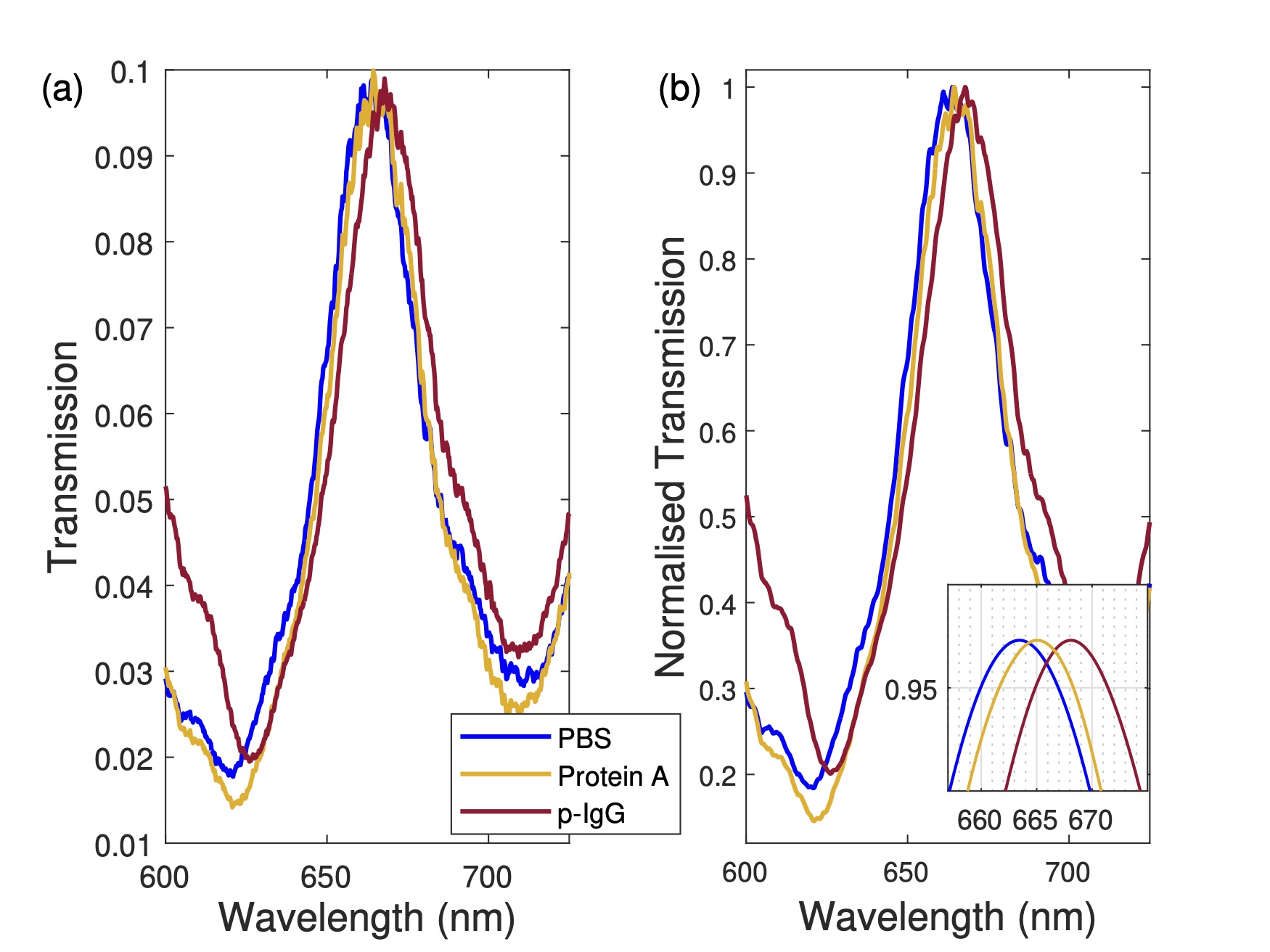}
\caption{Sensor functionalization measurements. (a) Measured spectra during different stages of the sensor functionalization. First, the response of the bare non-functionalized sensor in PBS was measured. Then, the Al-NHA was coated with 3-GPS and incubated in protein~A. After incubation, the NHA was washed in PBS, and measurements were taken. Finally, the Al-NHA was coated with plasmodium antibodies IgG1 specific to pLDH. After coating, the sensor was rinsed with PBS, and measurements were taken. (b) Normalized transmission spectra of (a). A moving mean algorithm was used to smooth the noise in (b). The inset figure is the spline fitting of the normalized spectra in (b). The legends presented in (a) are shared with Fig. (b).}
\label{fig6} 
\end{figure}
\begin{figure*}[t!]
\centering
\includegraphics[width=7cm]{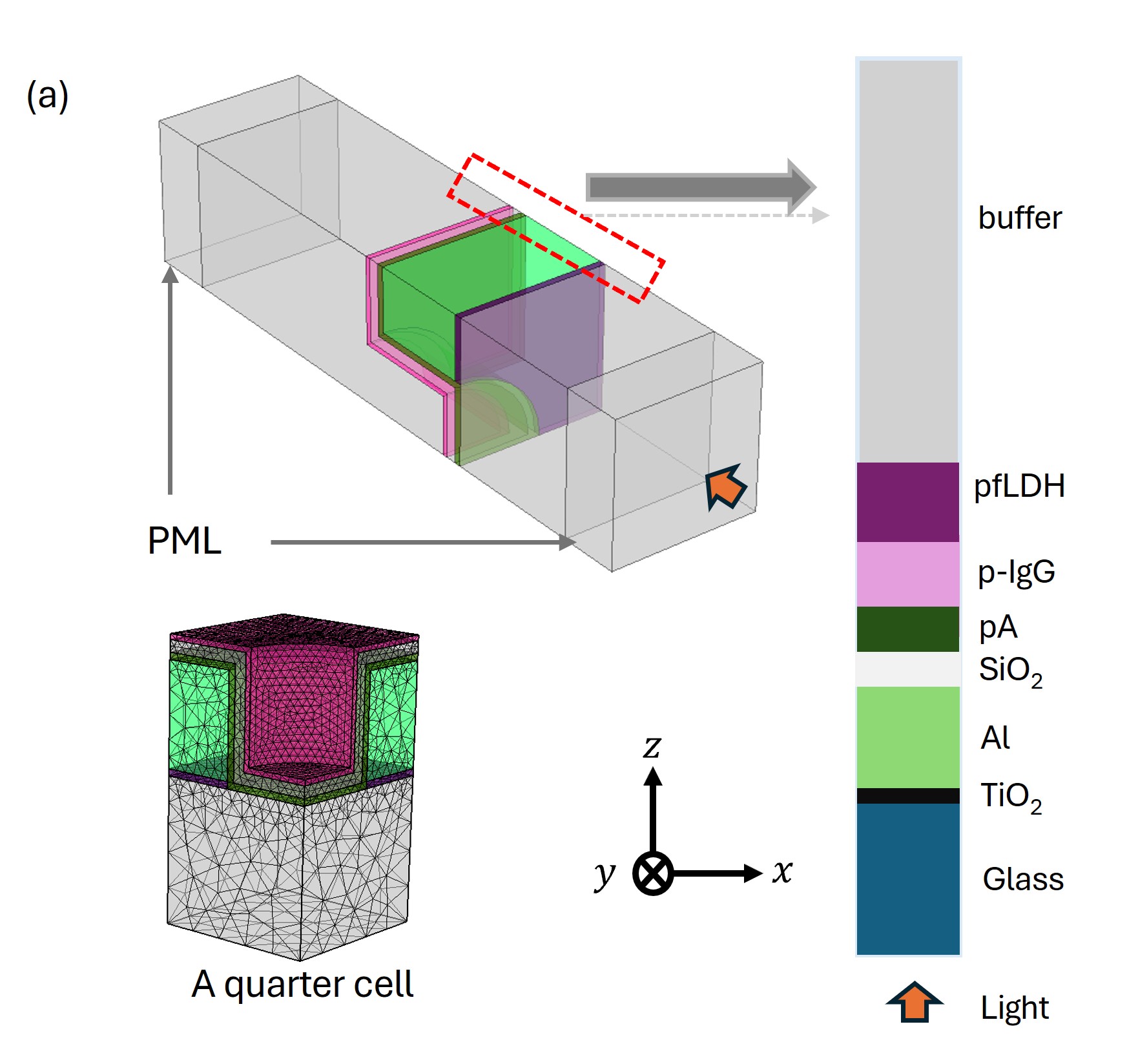}
\includegraphics[width=9cm]{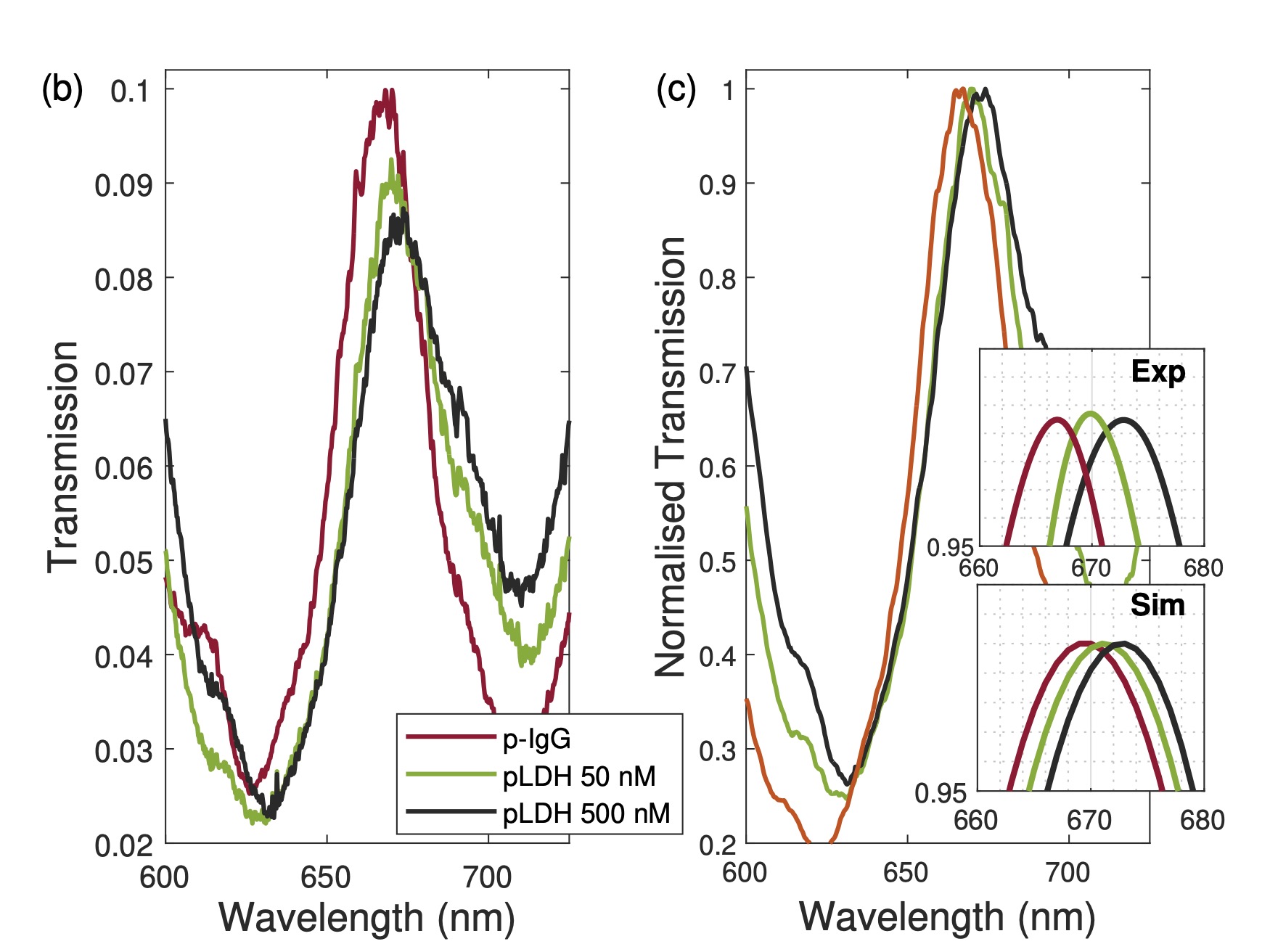}
\caption{Surface sensing of pfLDH. (a) The simulated sensor model design. The first and last domains are perfectly matched layers (PML). The bottom PML layer is followed by the glass substrate and an aluminum nanohole (a quarter cell). A TiO$_2$ layer is included as the adhesion layer between glass and aluminum. A 3-GPS layer is added on the SiO$_2$ surface, followed by a protein A layer (pA), the antibody layer (p-IgG), the analyte layer (pfLDH), and the buffer layer. User-controlled meshing was applied and optimized according to the geometrical features of each domain. The permittivity of Al and TiO$_2$ were taken from Ref.~\cite{rakic1998optical}. (b) Experimental spectra when $50$~nM and $500$~nM pfLDH solutions were added to the sensor. (c) The normalized spectra and a zoomed-in plot shows the spectral shift of pfLDH peaks from the p-IgG peak. The inset shows simulated peaks (Sim) that can be compared to the measured peaks (Exp). The legends in (b) are shared with (c) and the insets in (c).}
\label{fig7} 
\end{figure*}

Bulk sensing measures the sensor's response to changes in the bulk refractive index of the surrounding medium ($\textit{e.g}$. buffer solutions or solvents with known refractive index $n_{\text{eff}}$). This stage is known as the bulk sensing calibration stage, and it forms a baseline for later pfLDH-specific measurements. We consider how the non-functionalized sensor responds to a bulk change in the refractive index $n_{\text{eff}}$. We injected two organic solvents into the wet chamber, ethanol ($n_{\text{eff}} = 1.362$) and isopropanol ($n_{\text{eff}} = 1.377$), one at a time. The sensor was completely rinsed with PBS and air-dried before injecting another liquid. The corresponding transmission profiles (for $\lambda = 600$~nm to $725$~nm) are presented in Fig.~\ref{fig5}~(a). We measured the spectral resonance shift of EOT peak (ii), $\delta\lambda_R = \lambda_R^{\text{eth/iso}} - \lambda_R^{\text{pbs}}$, and calculated the sensor sensitivity $S_{\lambda_R}$, as defined in Eq.~\eqref{eq:Sens}. Experimental spectra are compared with the simulation spectra in the inset of Fig.~\ref{fig5}~(a). More details on the FEM simulation are given in Appendix~B. 

Sensitivity calculations are presented in Fig.~\ref{fig5}~(b). The sensor has a bulk sensitivity of $S_{\text{exp}} = 360$~nm/RIU, which is close to the simulated sensitivity of $S_{\text{sim}} = 355$~nm/RIU. These values are within the range of other aluminum~\cite{Victor2014, Lee2017} and gold-based EOT sensors~\cite{Ding2015}. For intensity sensing, we fixed the wavelength at $\lambda_0 = 660$~nm and plotted the sensor transmittance drop, $\delta T_{\rm norm}$, against the solvent refractive index to obtain the intensity sensitivity, $S_T$. The results are presented in Fig.~\ref{fig5}~(c). There is again a close match between the experimental ($S_{\text{exp}}=9.8876$~a.u./RIU) and~simulation sensor sensitivity ($S_{\text{sim}} = 9.2226$~a.u./RIU). 

In a practical clinical setting, bulk calibration would need to be done with whole blood lysate. Blood has a refractive index range that depends on the physiological condition of the patient and the temperature of the experiment. For example, in the different stages of malaria, blood has different refractive index values, from normal stage (no infection) at $1.408$, to ring stage at $1.396$, trophozoite stage at $1.381$, and schizont stage at $1.371$~\cite{park2008refractive, agnero2019malaria}. Once the blood is lysed in preparation for sensing, the values drop closer to that of PBS, in the range $1.335-1.35$. The specific refractive index value in this range affects the baseline position of the transmission peak used for sensing. As can be seen from Fig.~\ref{fig5}~(b) and (c), the sensor response is linear over a wide range of refractive index ($1.335-1.38$) for both wavelength and intensity sensing. Therefore the use of a blood sample should in principle cause only a small linear shift in the baseline position of either $\lambda_R$ or $T$ after the initial deposit of the sample on the sensor surface. If the sensor is functionalized (see next section) the transmission peak will shift further over a short period of time due to binding of pfLDH to the immobilised antibodies on the surface (if the blood is infected). This shift will be relative to the initial one and so the amount will not depend on the initial offset due to the state of the blood. 

\subsection{Sensor response: surface sensing}
The surface chemistry for the functionalization of the sensor has already been described in Section III A and B, in addition to Fig.~\ref{fig1}(d) and Appendix~C. Here, we describe the results of measurements taken during the steps outlined in Fig.~\ref{fig1}~(d). Between each step, the Al-NHA was washed and incubated in PBS. Measured spectra of the whole functionalization process are shown in Fig.~\ref{fig6}. Each spectrum is an average of 10 measurements. The EOT peak (ii) of the non-functionalized sensor in PBS was measured and found to be $\lambda = 664$~nm. The peak shifted to $\lambda = 665$~nm when protein A was attached to the 3-GPS layer. The final peak shifted to $\lambda = 668$~nm when p-IgG antibodies were immobilized. A total spectral shift of $4$~nm is therefore obtained due to functionalization. Our FEM simulation using the model presented in Fig.~\ref{fig7}~(a) and described next, gives a comparable total spectral shift of~$6$~nm.

\begin{figure*}[t]
\centering
\includegraphics[width=8.5cm]{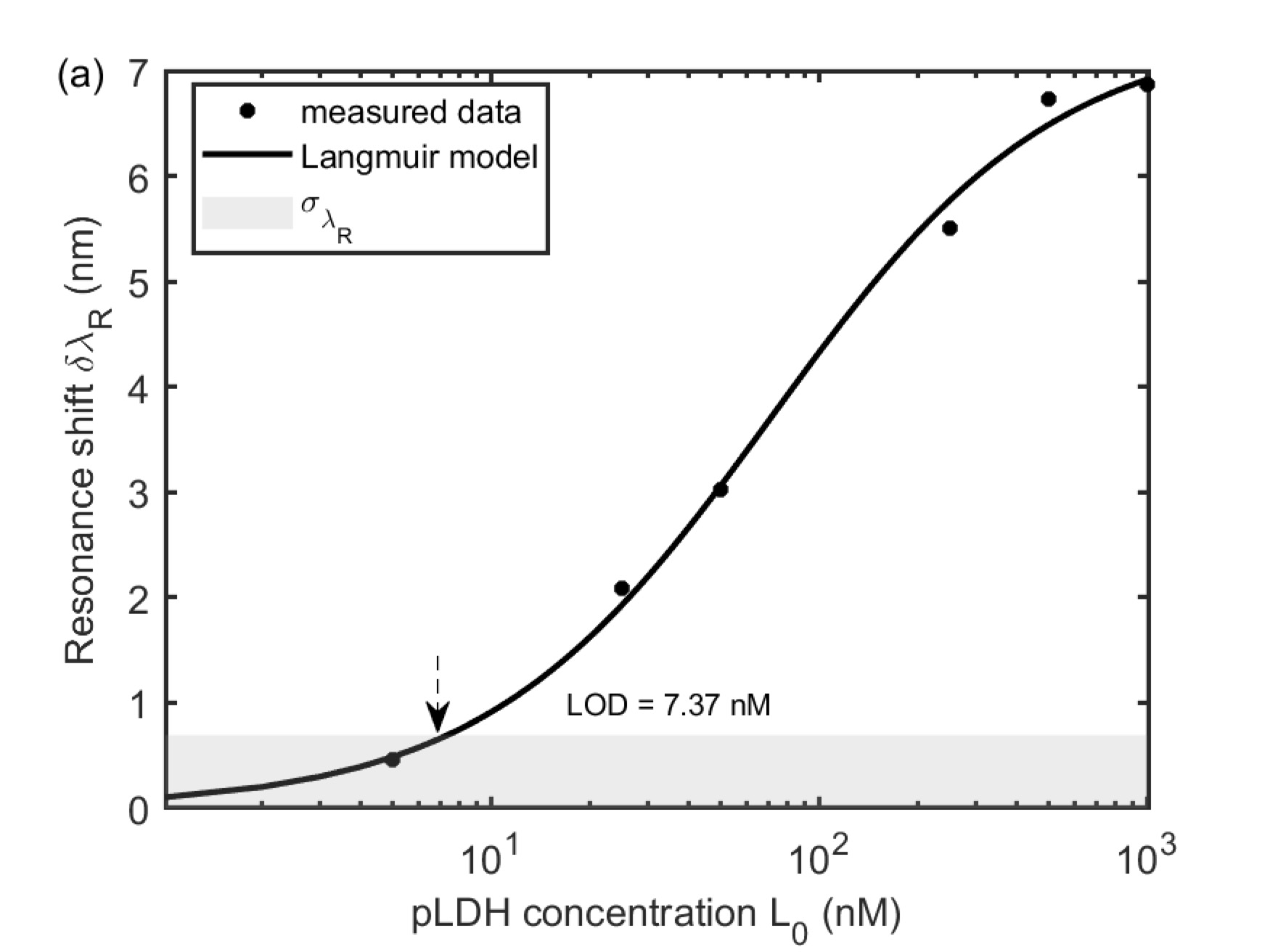}
\includegraphics[width=8.5cm]{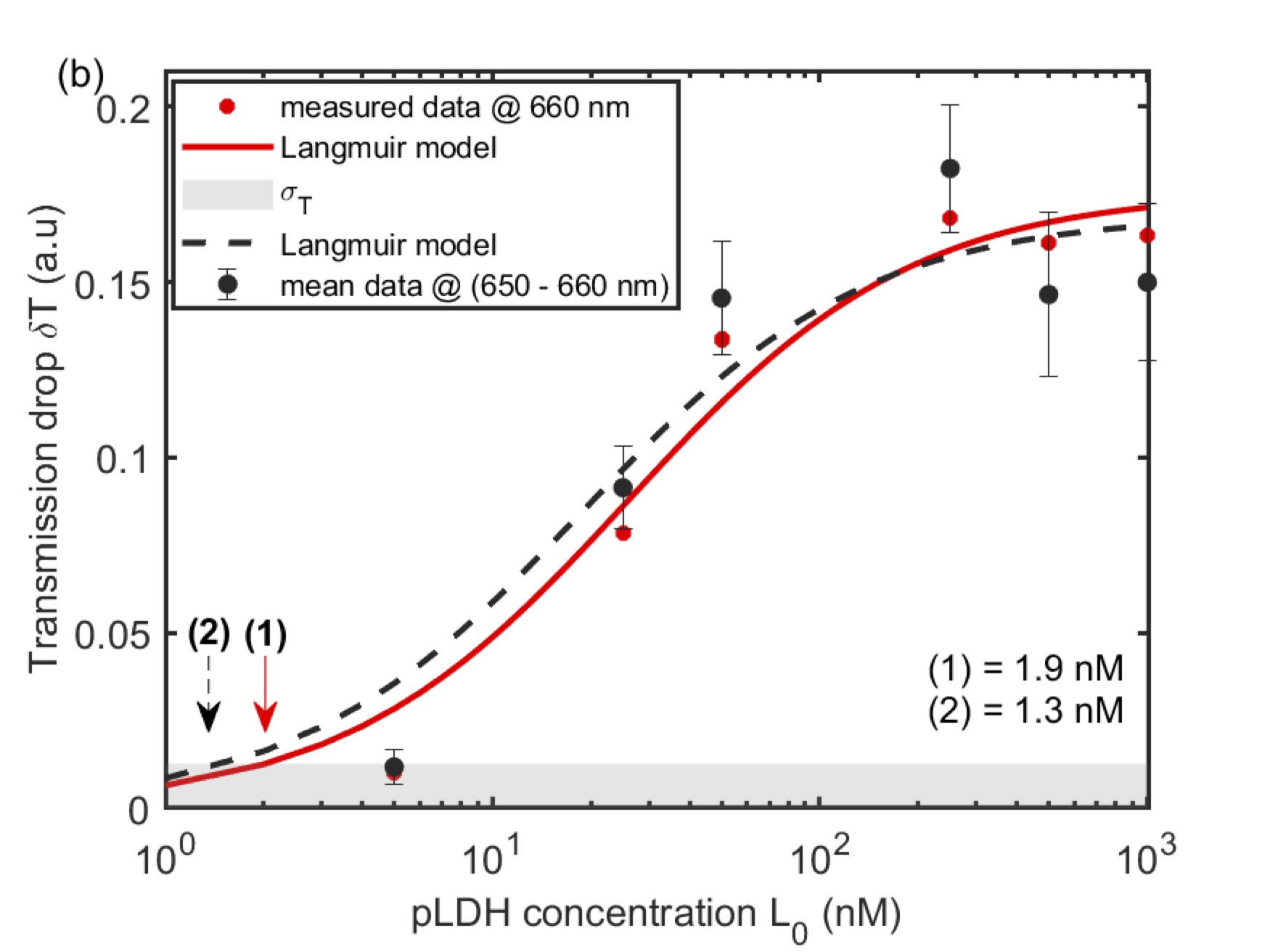}
\caption{Calibration curves and Langmuir model fitting. (a) Measured spectral peak shift due to pfLDH binding to immobilized antibodies. Concentrations ranging from $0 - 10^3$~nM are injected into the wet chamber. A limit of detection (LOD) of $7.37$~nM is obtained. (b) Normalized transmission drop measured at fixed wavelengths. Measured data (red points) and corresponding Langmuir model (red solid line) fitted for $\lambda = 660$~nm are displayed. Black points (measured data) with error bars represent average intensity sensing from 650~nm to 660~nm. The corresponding Langmuir model (black dashed line) is shown. The estimated LOD of $1.9$~nM for intensity sensing off-resonance at $\lambda = 660$~nm is shown by a red arrow labeled (1). The mean Langmuir model suggests an LOD improvement from $1.9$~nM (equivalent to 66.5 ng/mL) to~$1.3$~nM~(equivalent to 45.6 ng/mL) when an optimal wavelength is chosen between $650 - 660$~nm, shown by a black arrow labeled (2).}
\label{fig8} 
\end{figure*}

In Fig.~\ref{fig7}~(a), we show how the sensor model used in the simulations was designed in COMSOL. The refractive index and thickness of each layer are the key parameters in the sensor design. We used electromagnetic frequency-domain wave analysis in COMSOL to study the optical response of the sensor. A quarter cell was designed due to the rotational invariance of both excitation (plane wave) and material system (circular nanohole) along the wave propagation direction, normal to the sensor surface. Perfect electric conductor (PEC) and perfect magnetic conductor (PMC) boundary conditions were used. The former imposes symmetry for the magnetic fields by setting to zero the tangential component of the electric fields at the boundary. The latter imposes symmetry for the electric fields, prohibiting the electric current from flowing into the boundary. An interior periodic port based on floquet periodicity conditions was used to set up a plane wave excitation at the lower glass boundary. Meshing of the sensor domains was user-controlled. A free triangular fine mesh was assigned to all boundaries, and a free tetrahedral finer mesh was assigned to the Al-NHA, SiO$_2$ and TiO$_2$ domains. The remaining domains were assigned a free tetrahedral fine mesh. The top and bottom domains were terminated by strong absorbing perfectly matched layers. The permittivity of aluminum was taken from Ref.~\cite{rakic1998optical}. Refs.~\cite{Ding2015, Wong2013} present a similar design of a quarter-cell model. More details about the sensor design and parameters ($\textit{e.g}$. layer thickness and the dependence of the refractive index of the pfLDH layer with its concentration in~the bulk analyte) can be found in our theoretical study~\cite{Kiyumbi2024}. 

In Fig.~\ref{fig7}~(b) and (c), we show two representative experimental spectra of the sensor response. When pfLDH solution is added to the wet chamber, pfLDH molecules bind to the immobilized antibodies and result in a redshift of the p-IgG peak. It can be seen that there is a close match between experiment and simulation. When a pfLDH solution of $50$~nM was injected into the wet chamber, a redshift of $2.8$~nm from the baseline p-IgG spectrum was observed ($\textit{i.e}$. from  $668$~nm to $\sim 670$~nm). Simulations show a similar trend; a redshift of 2.2~nm is obtained. A similar behavior was observed when a pfLDH solution of 500~nM was added to the wet chamber, as shown in Fig.~\ref{fig7}~(b) and (c). A regeneration process was performed between each measurement by rinsing the sensor with PBS many times until the p-IgG spectrum was recovered. A $2.5$M sodium chloride (NaCl) solution prepared in PBS was also used to ensure complete regeneration. 

\subsection{Estimation of binding affinity and LOD}
We now turn our attention to the extraction of the equilibrium binding parameter, $K_{eq}^A$ (binding affinity), and the sensor LOD. In many kinetic controlled interactions like in our system, the adsorption of the analyte is usually modeled using a Langmuir adsorption model~\cite{Ayawei2017}. The system should reach equilibrium at the final stage of interaction, where the net effect of the association and dissociation process is zero. The equation describing this dynamic equilibrium is written in terms of $\delta\lambda_R$ or $\delta T$, as~\cite{Ayawei2017}
\be
\delta\lambda_R = \frac{\delta\lambda_{R,\text{max}}L_0}{K_{eq}^D+L_0},
\label{eq:Lmodel3}
\ee
and 
\be
\delta T = \frac{\delta T_{\text{max}}L_0}{K_{eq}^D+L_0},
\label{eq:Lmodel4}
\ee
where $L_0$ is the pfLDH concentration in the buffer and $K_{eq}^D$ is the equilibrium dissociation constant, defined as $K_{eq}^D = 1/K_{eq}^A$ (for more details see Ref.~\cite{schasfoort2017handbookb}). The $K_{eq}^D$ value is obtained by fitting the Langmuir model, Eqs.~\eqref{eq:Lmodel3} and~\eqref{eq:Lmodel4}, in a ``sensor response vs concentration" semi-log plot to the experimental data. Finally, from the relation $K_{eq}^D = 1/K_{eq}^A$, we can estimate the affinity parameter of the pfLDH-(p-IgG) interaction. On the other hand, the LOD is defined as the~lowest analyte concentration $L_{0,\text{min}}=\Delta L_0$ that produces a measurable~sensor response $R$ above the noise level, $\sigma_R$, as discussed earlier (see Eq.~\eqref{eq:LOD}).

In Fig.~\ref{fig8}, we plot the sensor responses, $\delta\lambda_R$ and $\delta T$ against the pfLDH concentration $L_0$. A log scale is used on the $x-$axis. A Langmuir model is fitted to the measured data, as given in Eqs.~\eqref{eq:Lmodel3} and~\eqref{eq:Lmodel4}, and the LODs are estimated from the graph based on Eq.~\eqref{eq:LOD}. The intersection of the $\sigma_R$ gray region and the Langmuir line fits (solid and dashed lines) determines the value of $\Delta L_0$ for a particular sensing scheme, as described in section III C. Comparing Fig.~\ref{fig8}~(a) and (b), we see that intensity sensing offers a better LOD of $1.9$~nM at $\lambda = 660$~nm compared to spectral sensing, which offers an LOD of $7.37$~nM. Moreover, the LOD at $1.9$~nM, can be improved to $1.3$~nM by choosing a suitable fixed wavelength between 650~nm and 660~nm. The optimal wavelength is $\lambda_0 = 655$~nm. The equilibrium dissociation constant of the pfLDH-(p-IgG) interaction is extracted from the model fitting and we find $K_{eq}^D = 71$~nM. The $K_{eq}^D$ value is comparable with those reported in Ref.~\cite{kD_pLDH2011}. Therefore, $K_{eq}^A$ is $1.408\times10^7 $M$^{-1}$, which is a good binding affinity
for use in the diagnosis of falciparum malaria.

The above analysis suggests that the best LOD is achievable with the Al-NHA in a transmission setup using the intensity sensing scheme. Moreover, spectral sensing is also limited by the spectrometer resolution~\cite{curry2007analysis}. Our spectrometer was able to resolve $\delta\lambda \sim 0.2$~nm. To resolve a smaller $\delta\lambda$ on the order of $0.01$~nm, due to a low pfLDH concentration, $\textit{e.g}$. in the pM range, would require a high-resolution spectrometer that would add cost to the sensor, but is possible in principle. The calculated affinity represents a moderate to strong binding affinity, indicating that the pfLDH-(p-IgG) interaction in this study is stable, specific, and biologically relevant. Many natural biological interactions, such as antibody-antigen binding or enzyme-substrate interactions, have $K_{eq}^D$ values in the nanomolar range, as discussed in Ref.~\cite{schasfoort2017handbook}. Our measured LOD exceeds a previously reported LOD of an EOT sensor based on gold nanoholes~\cite{Bohdan2020}.

Understanding what the LOD of our sensor means in a real medical diagnostic setting requires clinical data analysis of the pfLDH antigen levels in individuals infected with malaria. Here, to have an idea of this, we focus on the parasite density of individuals attending hospitals (patients) in tropical endemic regions. The reported parasite density ranges from 5000 parasites ($0.1\%$ parasitemia) to 50 parasites ($0.001\%$ parasitemia) per microliter of blood~\cite{Moody2002}. In malaria-naive individuals, pf infections of 500 to 1,000 parasites/$\mu$L are taken as the mean parasitemia to show symptoms (fever, chills, headache, etc)~\cite{Fryauff2000}. For the asexual stage, the relative amount of pfLDH produced by pf parasites at $0.001 \%$ parasite density (equivalent to 50 parasites/$\mu$L of blood) is estimated to be 1.87 ng/mL~\cite{Martin2009}. The performance of most rapid pLDH-specific mRDTs is relatively low in measuring such a density. For example, the LOD for the best pLDH-mRDT is reported to be 20 - 50 ng/mL~\cite{Tan2022, Jimenez2017}. A comparison of the performance between different malaria diagnosis techniques can be found in Refs.~\cite{Kasetsirikul2016, Fitri2022}. The Al-NHA sensor studied here offers a label-free detection of pfLDH with a modest LOD of $1.30$~nM, equivalent to $45.6$ ng/mL, which is comparable to that of mRDTs. The sensor could therefore be used in endemic regions as a stable alternative diagnosis method, with the added potential for detecting different parasite species. 

\subsection{Control experiments}
To investigate the sensitivity and specificity of our sensor we used recombinant human L-lactate dehydrogenase A ($\textit{i.e}$. hLDHA, SAE0049, Sigma-Aldrich) prepared in PBS. The protein is found in almost all living cells and it is structurally and functionally similar to pfLDH. Both enzymes catalyze the conversion of lactate to pyruvate and vice versa~\cite{Khrapunov2021}. They share a significant degree of sequence and structural homology, but have distinct epitopes that allow selective antibody recognition. Since clinical samples would naturally contain hLDHA, we believe that this is a realistic interference control and that demonstrating minimal cross-reactivity with hLDHA directly supports clinical specificity. 
\begin{figure}[t!]
\centering
\includegraphics[width=9cm]{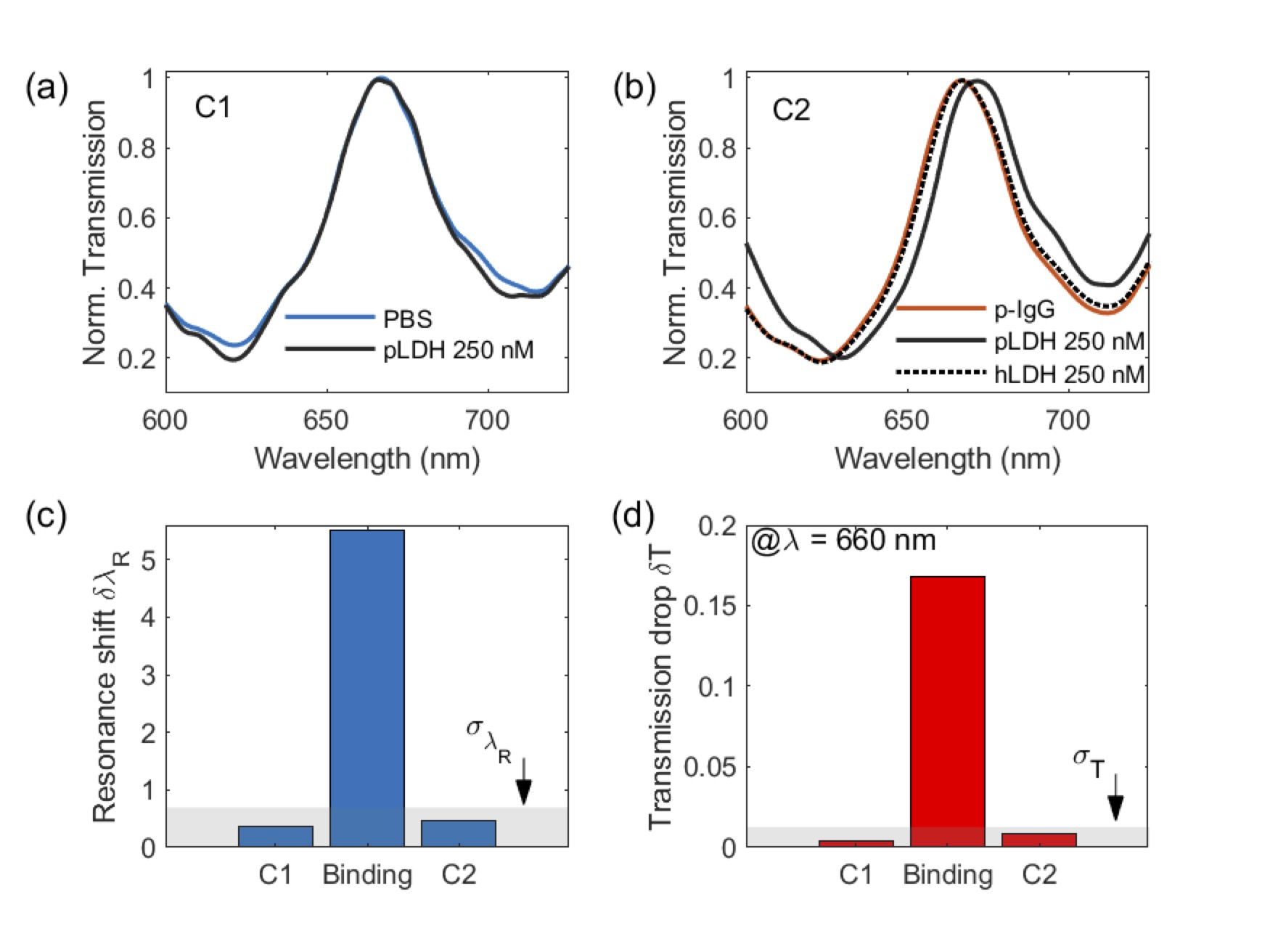}
\caption{Two control experiments. (a) The non-functionalized sensor response to pfLDH. In this experiment, C1, pfLDH 250~nM was added to the non-functionalized sensor. The sensor shows a near zero spectral shift from the baseline. (b) Human LDH (hLDH) and pfLDH of equal concentration were injected into a functionalized sensor. The sensor showed a low response below $\sigma_R$ when hLDH was added. (c) and (d) are bar graphs showing the spectral shift and transmission drops obtained in the two control experiments (a) and (b), respectively. Both C1 and C2 responses are below the sensor noise level $\sigma_R=3\Delta R$.}
\label{fig9} 
\end{figure}

In the first control experiment, labeled C1, and shown in Fig.~\ref{fig9} (a), 250~nM pfLDH solution was added to the non-functionalized sensor. The sensor showed a near-zero spectral shift from the baseline. In the second control experiment, labeled C2, and shown in Fig.~\ref{fig9} (b), human LDHA and pfLDH of equal concentration ($250$~nM), were added to a functionalized sensor, one after the other, with regeneration in between. The sensor responded well to pfLDH with a spectral shift of~$5.5$~nm, while with hLDHA, the sensor showed a low response below~$\sigma_R$, as shown in Fig.~\ref{fig9}~(c) as column C2. A similar behavior for the control experiments was observed for the transmission drop, as shown in Fig.~\ref{fig9} (d).

When the sensor is functionalized, the analytes are closer to the metal surface within the probe depth of the localized SP field. The signal response, $R$, of the sensor upon analyte adsorption to the immobilized receptor can be approximated as~\cite{Kiyumbi2024, Jung1998}
\be
\delta R = S_R(na_{\text{eff}} - nb_{\text{soln}})\left(1-e^{-2da/\delta_d}\right)e^{-2dc/\delta_d},
\label{eq:surf}
\ee
where $na_{\text{eff}}$ is the effective refractive index of the bound analyte layer, $nb_{\text{soln}}$ is the refractive index of the buffer, $da$ is the thickness of the analyte layer, $dc$ is the thickness of the biochemistry (3-GPS + Protein A + p-IgG) layer from the metal surface and $\delta_d$ is the probe depth of the localized SPs. For our sensor design, $S_{\lambda} = 360$~nm/RIU and $S_T = 9.8876$~a.u./RIU (see Fig.~\ref{fig5}), $na_{\text{eff}} = 1.45$, $nb_{\text{soln}} = 1.335$, $da = 8 $~nm, $dc = 19$~nm and $\delta_d$ ranges from $50-100$~nm. Using these values in Eq.~\eqref{eq:surf}, the expected spectral shift, $\delta\lambda_R$, of the sensor is $4.2 - 5.3$~nm and transmission shift, $\delta T$, is $0.14 - 0.18$ a.u. These shifts, $\delta \lambda_R$ and $\delta T$, are consistent with those observed in our experiment at a high concentration~of~$250$~nM, as shown in Fig.~\ref{fig9}(c)~and~(d), respectively.

Control experiment C2 shows that the immobilized p-IgG antibody is more specific to pfLDH than hLDH. For experiments that involve the detection of pfLDH in infected blood, a highly specific plasmodium antibody is important in order to distinguish between hLDH and pfLDH, and minimize significant interference from hLDH. A low affinity and specificity between pfLDH and the p-IgG antibody can lead to incorrect conclusions about the presence or severity of malaria because an increase in hLDH levels (which could occur due to various physiological conditions) might be mistakenly interpreted as an increase in the pfLDH level in affinity-based immunosensing. Therefore, further work on testing the sensor using whole blood lysate will be needed in this regard. 

Other proteins present in blood, such as HRP-2, or human immunoglobulins, such as HSA may also interact with the p-IgG antibodies, causing noise in the signal and impacting the sensor performance. A detailed study of non-specific adsorption and noise is a clear direction for future work. However, we believe the present study is an informative first step in this direction. 

Furthermore, interference effects from non-specific proteins can be addressed through surface functionalisation strategies~\cite{Lichtenberg19}. In general, such interference effects would impact the sensor by leading to a larger or smaller shift in the resonance peak position depending on the size and binding affinity of the protein. For instance, there may be a high binding affinity for a large non-specific protein. This would cause a large shift in the resonance peak compared to pfLDH and statistical fluctuations could mask the contribution of pfLDH to the shift. On the other hand, there may be a low binding affinity for a small non-specific protein, which would have only a small impact on the shift of the resonance peak. Various surface functionalisation strategies can be used to minimise false positives from binding of non-specific proteins. For example, one approach is to immobilise highly specific antibodies (or aptamers) on the sensor surface and ensure they have a uniform surface coverage during the immobilisation stage. A carefully deposited protein A/G layer can also help ensure that the antigen-binding sites of the antibody are accessible. One could also apply a blocking/anti-fouling layer to minimise non‐specific adsorption, for instance using HSA or casein. 

\subsection{Physiological concentrations of pLDH}
As mentioned earlier, pLDH concentrations in blood vary significantly across the infection stages and sampling time due to differences in parasitemia and parasite biomass. In early-stage infections, characterized by low parasitemia (typically $<\,200$ parasites$/\mu$L) and often asymptomatic or mild symptoms, pLDH levels are generally low and estimated at around $0.1-5$ ng/mL, based on correlations from sensitive immunoassays and patient samples~\cite{Martin2009, Moody2002, martianez2020quantification, jang2019simultaneous}. On the other hand, late-stage infections are characterized by elevated parasitemia levels (e.g., $>\,10,000$ parasites$/\mu$L), and the concentration of pLDH can increase substantially, reaching $150$ ng/mL or higher~\cite{jang2013pldh, Martin2009, Jimenez2017}. This escalation occurs as the enzyme is released during the maturation of the parasite and the rupture of schizonts. 

Our LOD of 1.3 nM (45.6 ng/mL) exceeds early-stage thresholds, however it reliably detects symptomatic ($1,000$ parasites$/\mu$L) and severe cases ($10,000$ parasites$/\mu$L), potentially supporting clinical use in endemic areas alongside more costly microscope imaging methods and PCR testing (which achieves an LOD of $\leq 0.1$ parasites/$\mu$L~\cite{mixson2010evaluation}). It should be mentioned that many infections (especially in endemic areas) are low-density (<100 parasites/$\mu$L) or even sub-microscopic (<10 parasites/$\mu$L), and these provide a reservoir for the disease to spread. More work is needed to improve the LOD of our sensor to achieve this level of detection and compete with microscope imaging methods and PCR testing. Furthermore, additional work on testing the sensor using whole blood lysate is needed to assess whether the LOD is sufficient for clinical deployment. 

\subsection{Benchmarking against existing SPR malaria sensors}
Protein-based planar SPR assays have already found application in malaria diagnostics~\cite{Ragavan2018}. They have been used to detect heme -- a malaria biomarker, to an LOD of $2\;\mu$M (1.3$~\mu$g/mL)~\cite{briand2012novel}. Sikarwar et al.~\cite{sikarwar2014surface} reported an LOD of 0.517~nM when an SPR sensor was used to identify and characterize monoclonal and polyclonal antibodies of pf-HRP2. This LOD is equivalent to an LOD of 2.90 nM if pLDH was to be detected in the sample~\cite{Martin2009}. Our LOD of 1.3 nM is therefore an improvement by a factor of two compared to that of Sikarwar et al. Moreover, the sensor of Sikarwar et al. is limited to the pf-species in principle, unlike our sensor. Further testing of other species of pLDH is of course needed for our sensor. Recently, a complex graphene-enhanced copper SPR biosensor for detecting malaria DNA sequences has been reported with an LOD in pM~\cite{wu2020ultrasensitive}. On the other hand, Cho et al.~\cite{Cho2013} reported the first experimental detection of pLDH in whole-blood lysate using a nanostructured sensor made from a gold film with a periodic NHA. The measured spectral sensitivity was $378$~nm/RIU in the visible spectrum, although no LOD was reported. Recently, Lenyk et al.~\cite{Bohdan2020} proposed a dual transducer malaria aptasensor. The method combines electrochemical impedance spectroscopy and SPR biosensing techniques in a single unit. An LOD of 23.5~nM was obtained based on SPR sensing. In addition to SPR techniques, other photonic studies have reported exploring possible sensors for the detection of malaria, many of which can be found in Refs.~\cite{Ragavan2018, Krampa2020}.

\section{Summary and Outlook}
Malaria has been a global health problem for many decades, taking the lives of about half a million people every year. The most available techniques (microscopy, PCR test, mRDT, etc.) face practical challenges in most resource-limited settings. An effective malaria diagnosis strategy using cost-effective biosensors, such as optical biosensors, can complement large-scale vaccination and eradication efforts launched in endemic African regions. Moreover, malaria biomarkers such as pLDH and Aldolase, key enzymes in the glycolytic pathway of the parasite, are required to detect early blood-stage infections and non-falciparum infections.

Among the different types of optical biosensors, plasmonic sensors have shown experimentally the possibility of detecting low concentrations of malaria biomarkers, such as pLDH spiked in buffer solution, or whole blood lysate. The best achieved LOD so far ranges from fM to pM based on a plasmon-enhanced fluorescence immunosensor. However, the achieved LOD based on fluorescence requires expensive equipment to detect the low light level of the fluorescence tags. On the other hand, for the SPR setup, the majority of demonstrated malaria plasmonic biosensors have been fabricated using gold and target pfHRP-2 as the malaria biomarker. The excellent plasmonic properties of gold -- low optical losses in the visible and near-infrared regions and its high chemical stability are counteracted by its high cost, which limits large-scale commercialization. An attractive alternative to gold is aluminum. Aluminum is easy to fabricate and functionalize, and has properties that enable plasmon resonances in a broad optical band. Nano-engineered aluminum surfaces, such as the one studied in this work, have previously been proposed for use in various other biosensing applications~\cite{Victor2014, Lee2017, martin2014Al_fabrication}.

We have shown that a low LOD is achievable using an aluminum metasurface as a plasmonic malaria sensor. The LOD of 1.3~nM, achieved using the Al-NHA sensor, is better than the previously reported LOD of 23.5~nM achieved using a gold-NHA~\cite{Bohdan2020}. Our results may encourage the design of cost-effective malaria sensors based on aluminum plasmonics for detecting the pLDH protein in real blood samples~\cite{Piper1999, Makler1998}. Due to the similarity in amino acid sequences among plasmodium species, it has been shown that monoclonal antibodies that target pfLDH can also detect P. vivax and P. knowlesi~\cite{kD_pLDH2011}. Therefore, the sensor could potentially be used to detect any species' pLDH for biochemical research and point-of-care in the clinical field. 

Furthermore, the surface chemistry using an epoxy-silane linking protocol followed in this study is simple and effective. Both the Al-NHA and the surface chemistry developed are stable in liquid environments and can be used for a long time without degradation in sensor properties. Non-specific adsorption noise was minimized via protein A orientation and 3-GPS silanization, reducing false positives compared to direct IgG immobilization. 

On one hand, oriented antibody immobilization via protein A can be considered to be undesirable for sensitivity, as it causes an attenuation in the plasmonic field at the location of the biomarker (due to an increase in the distance of the biomarker from the metal surface), however this is balanced by substantial improvements in binding capacity and selectivity. Previous studies have reported over a 50\% enhancement in plasmonic sensing performance when using protein A, with net sensitivity improvements and a 1.5-fold increase in the detection limit~\cite{Chung2006,Coluccio2021}. While we did not explicitly check the impact of the protein A layer in our experiment, we expect a similar behavior to these studies. In general, the orientation of antibodies via an intermediate layer, such as protein A, is of crucial importance in many diagnostic devices for reducing the LOD and improving specificity. We evaluated the specificity of the IgG antibody using human LDH in our control experiments. Control experiment C2 shows that the immobilized p-IgG antibody is far more specific to pfLDH than hLDH (non-specific adsorption noise). 

For clinical validation, further tests are needed for studying other proteins present in blood that might also interact with the immobilized IgG antibodies, including a detailed study of non-specific adsorption in a real blood sample. The best case scenario is to use a low $K_D$ (high binding affinity) antibody~\cite{kD_pLDH2011,Kaushal2014}, aptamer~\cite{Minopoli2020,wu2020ultrasensitive,RoyeroBerme25} or other biorecognition chemistry to minimize non-specific binding, and improve selectivity and regeneration. Aptamers in particular offer excellent prospects for achieving high binding affinities leading to ultra-low LODs ($<30$~fM)~\cite{Minopoli2020,wu2020ultrasensitive}, although surface functionalization is more involved and expensive imaging equipment may be required. 

Another aspect of further study is the relation between the sensitivity and signal-to-noise (SNR) ratio of the sensor. In principle they are unrelated, however, for our sensor, the sensitivity depends on the NHA geometry and the spatial extent of the field in the sensing medium, while the SNR also depends  on the NHA geometry, as the signal value (ie., transmitted intensity) is determined by the transmission efficiency through the NHA, and a low signal value is more susceptible to other technical sources of noise. The relation between the sensitivity and SNR is therefore complicated and trade-offs need to be carefully assessed throughout the design, simulation and testing stages for any new version of the sensor. 

In the future, we also plan to test the sensor using intensity interrogation with a stable narrow bandwidth laser in the 650 - 670~nm range to enhance the overall sensitivity and reduce noise. In addition, the sensor's LOD may be improved by using more specialised antibodies with lower $K_D$ values~\cite{kD_pLDH2011,Kaushal2014}. Another direction to improve the LOD would be to replace the classical source of light with alternative quantum light sources~\cite{Lee2021} that could offer an improved LOD due to a decrease in shot noise~\cite{Mpofu2021}.

\section*{Acknowledgements}

This project was financially supported by the African Laser Centre (ALC) grant (ALC-R016) at the South African Council for Scientific and Industrial Research (CSIR). The authors thank Dr. Gurthwin Bosman, Daniel Retief, Zahra Tayob, and Dr. Luke Ugwuoke for their supportive ideas. We also acknowledge administration support from the University of Dar es Salaam, Dar es Salaam University College of Education, Tanzania, the Stellenbosch Photonics Institute (SPI), the Department for Science and Innovation (DSI), and the National Research Foundation (NRF) of the Republic of South Africa.

\setcounter{figure}{0}
\setcounter{subsection}{0}
\renewcommand{\thesubsection}{\Alph{subsection}}
\renewcommand\thefigure{\thesubsection.\arabic{figure}}  

\section*{Appendix}

\subsection{Sensor topography and dimension} 
\begin{figure}[b!]
\centering
\includegraphics[width=8cm]{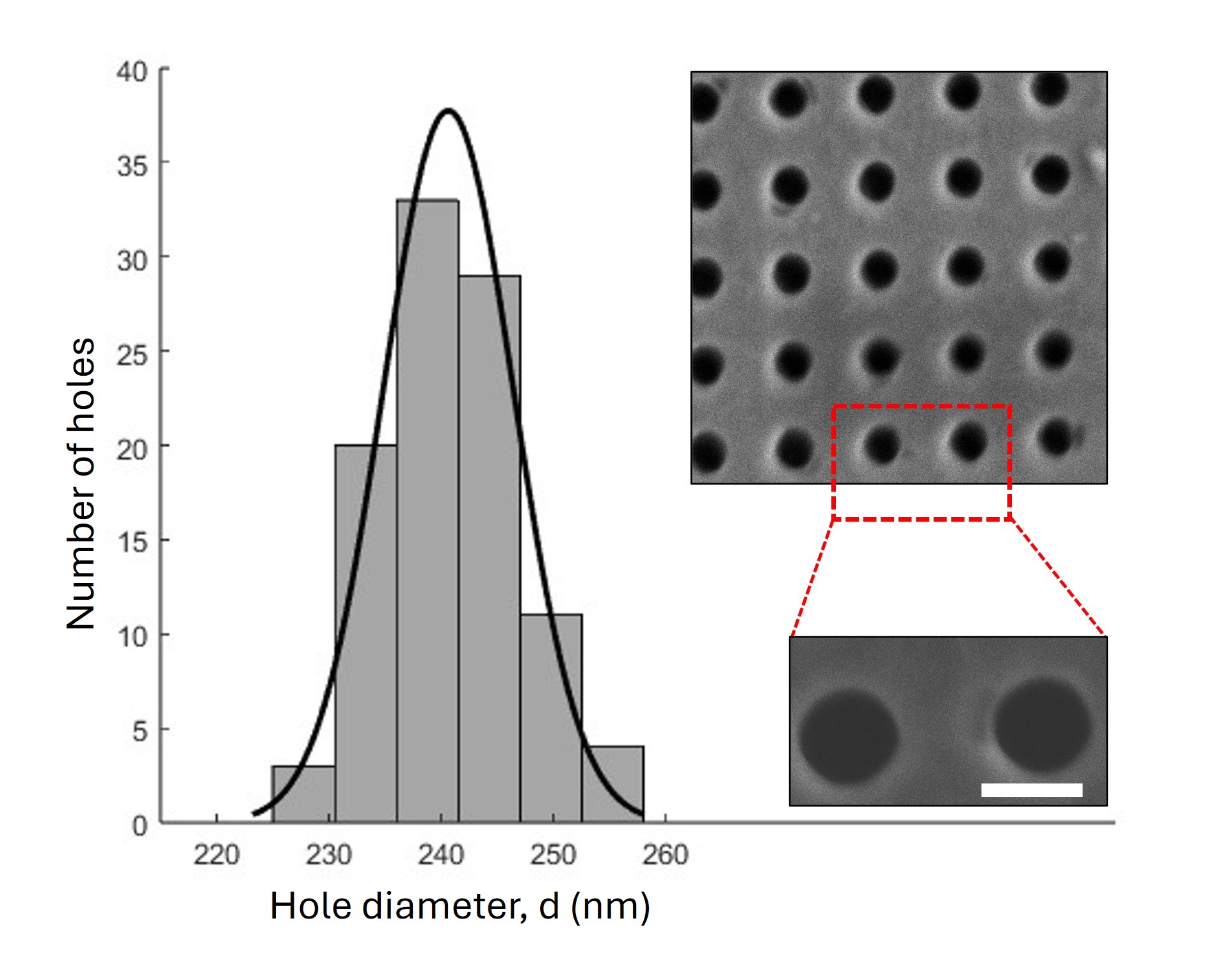}
\caption{Sensor topography and nanohole diameter $d$ distribution. The inset is the scanning electron microscope (SEM) image of the~Al-NHA~metasurface sensor. The scale (white bar) represents $250$~nm. Based on statistical analysis, $i.e.$ a Gaussian fit for the histogram, the~mean~hole diameter $d$ is $242$~nm and the NHA period $a_0$ is $450$~nm.}
\label{fig2} 
\end{figure}

The scanning electron microscope (SEM) image of our sensor is presented in Fig.~\ref{fig2} showing the sensor topography, hole distribution, and lateral geometric parameters. Using a Gaussian fit for the histogram, on average, each hole in the array has a diameter $d~=~242$ nm and period $a_0 = 450$ nm.

\subsection{Finite element method (FEM) simulation details}
\setcounter{figure}{0}
\begin{figure*}[t!]
\centering
{\includegraphics[width=0.85\textwidth]{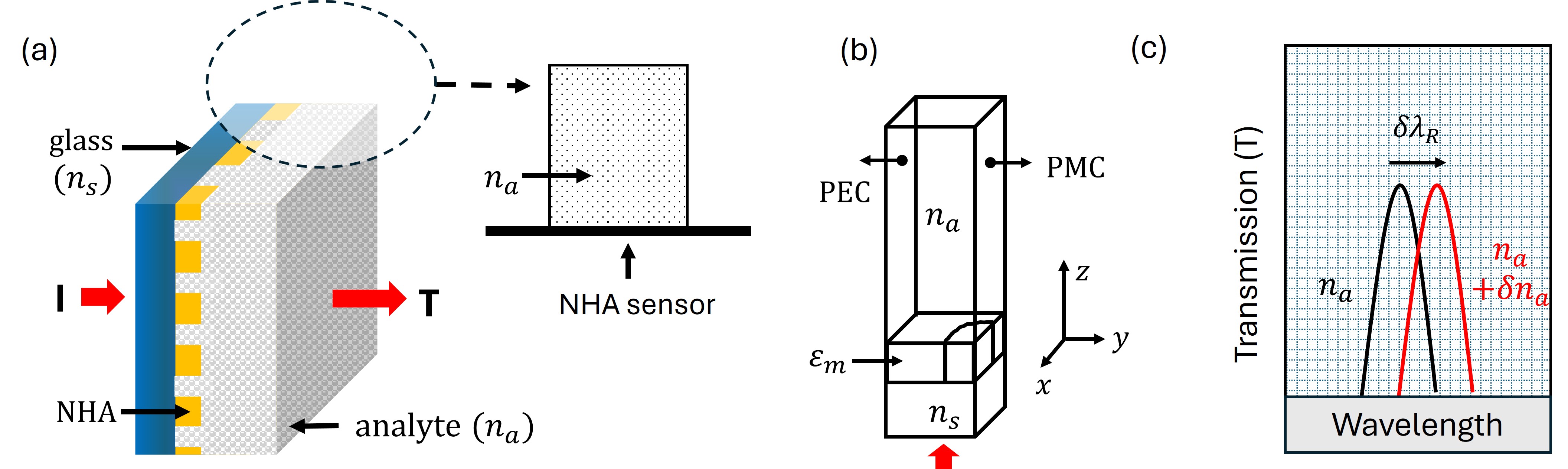}}
\label{non-func-sensor}
\caption[]{Finite element simulation of a non-functionalized sensor. (a) A schematic drawing of the Al-NHA sensor showing the three main components of the sensor, the substrate layer (glass), nanohole array (optical transducer), and the analyte layer (e.g. PBS). (b) A COMSOL model of the non-functionalized sensor showing a quarter of a unit cell. A TM-polarized light field--red arrow ($E$ along $y$, $H$ along $x$, and propagation along $z$) is incident on the quarter cell from the glass side. For a periodic NHA response, PEC and PMC boundary conditions are applied perpendicular to the $x$ and $y$ axes, $i.e$, $yz$ plane and $xz$ plane, respectively, ensuring the quarter cell accurately represents an infinite NHA. Top and bottom planes are port boundaries: The bottom plane ($z = 0$) allows incident and reflected waves, while the top plane allows transmitted waves. (c) The working principle of the sensor. When the refractive index $n_a$ of the analyte solution ($e.g.$ PBS) near the metal surface is altered ($e.g.$ from PBS to ethanol or isopropanol), $n_a \rightarrow n_a +\delta n_a$, and the EOT peak position changes. By monitoring the spectral change, $\delta\lambda_R$, for example, the bulk sensitivity, $\delta\lambda_R/\delta n_a$, can be obtained.}
\label{sensor_design}
\end{figure*}

\label{appendix_B}
The finite element method (FEM) simulations were performed using COMSOL Multiphysics (version 5.6). A 3D unit cell of the square lattice nanohole array (NHA) sensor shown in Fig.~\ref{sensor_design}(a) was modeled, with Floquet periodic boundary conditions applied in the $x$ and $y$ directions to represent an infinite array of periodic nanoholes. To leverage symmetry and reduce computational time, a quarter of the unit cell was simulated, as illustrated in Fig.~\ref{sensor_design}(b). The boundaries perpendicular to the x-axis were set as perfect electric conductors (PEC), and those perpendicular to the y-axis as perfect magnetic conductors (PMC). The PEC and PMC are symmetry-enforcing boundary conditions and used in the quarter-cell COMSOL model to replicate the behavior of an infinite periodic array while reducing computational load. 

The top domain was defined as the analyte layer (refractive index $n_a = 1.335$ for PBS), the bottom as the glass substrate ($n_s = 1.52$), and the middle as the aluminum film perforated with sub-wavelength holes (thickness $t \approx 150$ nm, period $a_0 = 450$ nm, hole diameter $d \approx 250$ nm). The frequency-dependent permittivity of aluminum ($\varepsilon_m$) was incorporated using the Drude-Lorentz model from Rakic et al.~\cite{rakic1998optical}. A normal-incidence TM-polarized plane wave was used as the optical excitation from the substrate side, with wavelengths scanned from $450-900$ nm. To prevent numerical reflections, perfectly matched layers (PML) terminated the top and bottom domains. The mesh consisted of tetrahedral elements with a maximum size of $\lambda/10$ (where $\lambda$ is the wavelength) near the metal-dielectric interfaces for numerical accuracy. Free tetrahedral meshing is used for domains, with free triangular meshing for periodic boundaries. A fine mesh balances computational efficiency and accuracy. Electric field distributions were computed at key resonance peaks (e.g., 545~nm in air, 665~nm in PBS, 763~nm at the interband transition) by solving the time-harmonic Maxwell's equations using FEM, yielding the spatial distribution of the electric field vector, and normalized as $|E|/|E_0|$, where $E_0$ is the incident field amplitude. 

\subsection{Sensor functionalization}
\setcounter{figure}{0}
\begin{figure*}[t!]
\centering
\includegraphics[width=16.5cm]{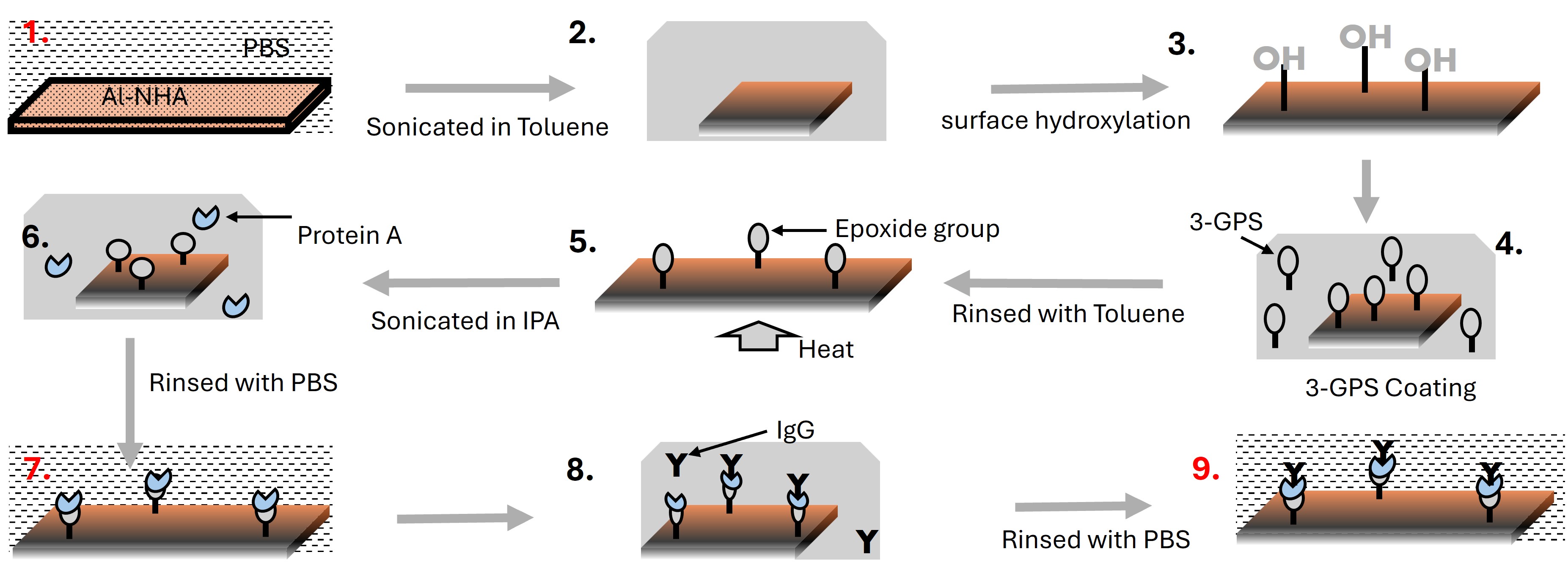}
\caption{Sensor functionalization protocol. (1) The  SiO$_2$ passivated Al-NHA in PBS. (2) The Al-NHA is cleaned in toluene to remove organic contaminants. (3) O$_2$ plasma treatment to increase the abundance of hydroxyl group -OH. (4) 3-GPS incubation. The Al-NHA was incubated in $1\%$ concentration of 3-GPS in toluene to allow the conversion of a silanol-terminated surface to an alkylsiloxy-terminated surface, $\textit{i.e}$. epoxide terminated surface. (5) The Al-NHA is rinsed in toluene and placed in an oven at $130^{\circ}$ C. (6) Protein A is attached to the surface via an amine bond. (7) The epoxide group reacts with amine nucleophiles of protein A. (8) The sensor is rinsed with PBS, and p-IgG solution is added. (9) Protein A binds the Fc regions of the p-IgG, exposing the Fab binding sites for the pLDH protein to bind. Red numbers represent stages where spectra were measured and the corresponding spectra (normalized to the white light spectrum) are shown in Fig.~\ref{fig6} of the main text.}
\label{appendex2} 
\end{figure*}
In Fig.~\ref{fig1}~(d) (main text) and Fig.~\ref{appendex2}, we outline the protocol for the sensor functionalization process. The process is discussed in more detail in Refs.~\cite{makaraviciute2013site, Cho2013}. The Al-NHA is passivated by a SiO$_2$ layer to prevent the surface from oxidation and corrosion. The passivated Al-NHA has a silanol-terminated surface. Proteins like immunoglobin (IgG) are unable to form bonds with the silanol-terminated surface without a silane coupling agent~\cite{Hermanson2013chap13}. Therefore, (3-Glycidoxypropyl)trimethoxysilane (3-GPS), a silane coupling agent that can link the inorganic silica surface with antibodies ($\textit{e.g}$. p-IgG) directly, or via protein A~\cite{Hermanson2013chap13, Hermanson2013chap15, makaraviciute2013site, Yang2003} was used in this study. It is a highly versatile dual-functional molecule containing an epoxy ring reactive group on one side and silanol functional groups on the other side~\cite{Hermanson2013chap13}. These properties enable it to chemically bond to the inorganic surface, which contains the -OH group, and organic/ligand biomaterials ($\textit{e.g.}$ protein A) that contain the thiol, amine, or hydroxyl group~\cite{Hermanson2013chap15, Nam2006, silberzan1991silanation, Yang2003}. The 3-GPS supplied by Sigma-Aldrich with $\ge 98\%$ purity was used without further purification. 

The Al-NHA is first cleaned in toluene and hydroxylated (O$_2$-plasma treatment) to ensure an abundance of reactive~-OH groups on the silica surface. The hydroxylated Al-NHA is then incubated for 30 minutes in 1\% (v/v) 3-GPS solution prepared in toluene. Surface silanol groups of the SiO$_2$ layer react with the 3-GPS silane (alkoxy
or chlorine) group, forming a Si-O-Si bond and releasing methanol as a byproduct~\cite{Hermanson2013chap13, Hermanson2013chap15}. This process is called silanization, $\textit{i.e}$. the conversion of a silanol-terminated surface to an epoxy-terminated surface~\cite{Nam2006, silberzan1991silanation}. The silanization process was enhanced by applying heat ($130^{\circ}$C) that accelerates the formation of the Si-O-Si bonds and helps in the removal of the methanol by-product. 3-GPS monolayer coatings ($0.8-1.2$~nm~thickness) are stable in liquid environments because the epoxy group provides better protection against hydrolysis and degradation~\cite{Hermanson2013chap13, Hermanson2013chap15}.

The Si-O-Si bond anchors the 3-GPS molecules onto the silica surface. The epoxy ring at the end of the 3-GPS molecule provides stronger covalent bonds when attaching an amine-containing biomolecule like protein A. The pfLDH antibodies (p-IgG) conjugation protocol followed in this study was adapted from Ref.~\cite{Parkkila2022, Cho2013}. It involves: (i) Protein A attachment to the sensor surface. Protein A is anchored to the sensor through amine bonding with the epoxy ring of the 3-GPS. (ii) The p-IgG antibodies are then immobilized on the sensor through protein A. The p-IgG antibody solution was prepared from a mouse anti-plasmodium species LDH (mouse IgG1, clone M301, from the Native Antigen Company), $\>90\%$ pure by SDS-PAGE and was presented in phosphate-buffered saline, pH 7.3 with $0.05\%$ sodium azide. The protein A-coated-Al-NHA was incubated in a $50$ $\mu$g/mL p-IgG solution prepared in PBS for 4 hours at room temperature. More details about the~protocol applied in this work are given in Refs.~\cite{Hermanson2013chap13, Hermanson2013chap15}.

\bibliography{myreferences} 
\end{document}